# Spatiotemporal fluctuation scaling law and metapopulation modeling of the novel coronavirus (COVID-19) and SARS outbreaks

*Zhanshan (Sam) Ma*
Computational Biology and Medical Ecology Lab
State Key Laboratory of Genetic Resources and Evolution
Kunming Institute of Zoology
Chinese Academy of Sciences, Kunming, China
&
Center for Excellence in Animal Evolution and Genetics
Chinese Academy of Sciences, Kunming, China
Email: ma@vandals.uidaho.edu

## Abstract

We comparatively analyzed the spatiotemporal fluctuations of the 2019-novel coronavirus (COVID-19) and SARS outbreaks to understand their epidemiological characteristics. Methodologically, we introduced TPL (Taylor's power law) to characterize their spatiotemporal heterogeneity/stability and Hubbell's (2001) unified neutral theory of biodiversity (UNTB) [specifically Harris *et al*. (2015) HDP-MSN model (hierarchical Dirichlet process—multi-site neutral)] to approximate the metapopulation of coronavirus infections. First, TPL analysis suggested that the coronaviruses appear to have a specific heterogeneity/stability scaling parameter (TPL-*b*) slightly exceeding *2* for cumulative infections or exceeding *1* for daily incremental infections, suggesting their potentially 'chaotic', unstable outbreaks. Another TPL parameter ($M_0$) (*i.e*., infection critical threshold) depends on virus kinds (COVID-19/SARS), time (disease-stages), space (regions) and public-health interventions (*e.g.,* quarantines and mobility control). $M_0$ measures the infection level, at which infections are random (Poisson distribution) and below which infections follow uniform distribution and may die off if $M_0$ coincides or below the level of Allee effects. For example, $M_0$=5.758 (COVID-19, China, or ***196***=$M_0$x34 in total) *vs*. $M_0$=2.701 (COVID-19, World) *vs*. $M_0$=9.475 (SARS, World) suggested that the potentially 'stabilized' infection level of COVID-19 in China is nearly ½ lower than that of SARS worldwide, but twice higher than that of COVID-19 worldwide. This may indicate that COVID-19 outbreak seems nearly twice more risky than SARS, and the lower infection threshold may be due to its lower lethality than SARS since lower fatality rates can facilitate the survival and spread of pathogen. Second, metacommunity UNTB neutrality testing seems appropriate for approximating metapopulation of coronavirus infections. Specifically, two parameters *θ* and *M,* borrowed from neutral theory, may be used to assess the relative significance of "infection through local contagion" *vs*. "infection through migration", both of which may depend on time, space, virus kinds, and particularly public-health interventions. Third, comparing the levels of TPL-$M_0$ and $\theta$ may leverage the complementary nature of TPL analysis and metapopulation modeling. For example, their difference ($\theta$–$M_0$=0.6) for the COVID-19 in China confirmed the critical importance of controlling migration (mobility) in suppressing the outbreak, although migration may have been a primary driving force for the initial outbreaks in several regions of China.

## Introduction

The spatial and/or temporal distributions of many biological populations including microbes and humans follows Taylor's power law (TPL) (Taylor 1961, Taylor *et al.* 1977, 1983, 1988), and recent studies have also revealed its applicability at the community scale (Ma 2012a, 2015, Li & Ma 2019, Taylor 2019, Ma & Taylor 2020). TPL has been verified by hundreds if not thousands of field observations in nature (Taylor 2019), and it has also found cross-disciplinary applications beyond its original domains of biology and ecology in disciplines such as computer science, natural disaster modeling, and experimental physics (Eisler et al. 2008, Ma 2012b, 2013, Tippett & Cohen 2016, Helmrich *et al*. 2020). Despite that TPL was proposed more than a half century ago (Taylor 1961) and there is not yet a consensus on the underlying mechanisms generating TPL (Eisler 2008, Stumpf & Porter 2012), there seems to be a recent resurgence of interests in this near universal model that captures the relationship between the population mean ($m$) and variance ($V$) in the form of a simple power function (*i.e.*, $V=am^b$) (*e.g.*, Cohen & Schuster 2012, Cohen & Xu 2015, Giometto et al. 2015, Kalinin *et al*. 2018, Oh et al. 2016, Reuman et al. 2017). Among the numerous existing studies on TPL, there have already been its applications to the analyses of spatial variation of human population (Cohen *et al.* 2013), human mortality (Bohk *et al*. 2015), and epidemiology (Rhodes & Anderson 1996). Given these previous applications to human demography and epidemiology, we postulate that TPL should also be applicable to the outbreak analyses of coronavirus infection diseases such as the still ongoing 2019 novel coronavirus (2019-nCoV) or COVID-19 (coronavirus-infected Pneumonia disease) (https://www.who.int/emergencies/diseases/novel-coronavirus-2019) (Li *et al.* 2020,

Thompson *et al*. 2020, Kucharski *et al*. 2020, Zhang *et al*. 2020) and 2003 SARS (Severe Acute Respiratory Syndrome) (https://www.who.int/csr/sars). In the present report, we test this hypothesis and further explore possible epidemiological processes (mechanisms) underlying the outbreak of COVID-19 infections.

While TPL can be harnessed to investigate the spatiotemporal fluctuations of coronaviruses, specifically, the scaling (changes) law of coronaviruses infections over space and time, we also aim to understand the spread of the virus infections from both local contagion (endemic) and external migration (epidemic and pandemic) perspectives. Nevertheless, this can be rather challenging given the lack of controlled experimental data, which is ethically infeasible to collect obviously. In principle, all of the infections existing globally constitute a metapopulation of people infected by the coronavirus, but constructing standard epidemiological models (*e.g.*, Wang et al. 2018, Rivers *et al.* 2019) with existing data is rather difficult. We realized that Hubbell's (2001) neutral theory of biodiversity, which is one of the four major metacommunity models (the other three include species sorting, mass effect, and patch dynamics) (Rosindell *et al*. 2011, 2012; Vellend 2010, 2016), might be adapted to approximate the meta-population dynamics. This approximation allows us to obtain, to the minimum, an educated guess for the local contagion spread and global dispersal (migration) parameters of the coronavirus infections.

Overall, this study sets two primary objectives: (*i*) to investigate the spatiotemporal fluctuation scaling law and (*ii*) to obtain an educated guess for the local contagion spread and global migration parameters of the COVID-19 infections. In addition, we also perform comparative analyses with the SARS to get more general insights on the epidemiology of coronavirus infections. To the best of our knowledge, this should be the first systematic application of TPL and UNTB in epidemiology, and obtained scaling/contagion/migration parameters should also be of significant biomedical importance.

## Datasets and Methods

**Datasets of COVID-19 and SARS infections**

We collected the worldwide, daily incremental and cumulative infections of 2019 novel coronavirus (COVID-19) and SARS, respectively. For the datasets collected in China, the unit of data collections was set to Chinese provinces. In addition, for the COVID-19 infections, we also collected the datasets of 17 cities of Hubei province of China. For the worldwide COVID-19 infections, the unit of data collections was set to country or region recognized by the WHO (world health organization). The date range for collecting the SARS data was between March 17 and August 7 of 2003 (136 days), and that for COVID-19 was between January 19 and Feb 29, 2020 (40 Days). Since the COVID-19 infections are still continuing, the analyses conducted in this report may be updated periodically.

**TPL (Taylor's power law) fluctuation scaling law for the infections of coronaviruses**

Taylor (1961) discovered that the relationship between mean abundance (*m*) and corresponding variance (*V*) of biological populations follows the following power function,

$$V = am^b, \qquad (1)$$

where *b* is termed *population aggregation* parameter and is thought to be species-specific, and *a* is initially thought to be related to sampling schemes used to obtain the data. The relationship is known as Taylor's power law (TPL) in literature, and it has been validated by hundreds, if not thousands of field observations worldwide (Taylor 2019). With a simple log-transformation, TPL can be converted into the following log-linear model:

$$\ln(V) = \ln(a) + b\ln(m). \qquad (2)$$

TPL was initially discovered in fitting the spatial or cross-sectional sampling data (Taylor 1961) and later found that it is equally applicable to temporal or time-series sampling data (Taylor 2019). In the context of time-series modeling, *b* measures the *population stability* (*variability*). More recently, it was found that TPL can be extended to community level from its original population level (Ma 2015). At the community level, the four Taylor's power law extensions (TPLE), can be used to measure the community spatial (temporal) heterogeneity (stability) (Type I & II), as well as mixed-species level spatial (temporal) heterogeneity (stability) (Type III & IV) (Ma 2015, Li & Ma 2019, Ma & Taylor 2020). Note that the term *aggregation* at the population level can be considered as the counterpart of *heterogeneity* at the community level.

In general, there are two important aspects related to the applications of TPL. First, test the fitting of the TPL model to the datasets under evaluations (such as SARS or COVID-19 infections). The testing determines the applicability of TPL based on well-established statistics such as *p*-value or *R* (linear correlation coefficient). Second, interpret the TPL parameters based on both the general principle of TPL (explained above) and system- or data-specific information (such as the biology of COVID or SARS).

Specifically, regarding the first aspect or the fitting of TPL, we adopt two fitting approaches: one is the simple linear regression *via* log-transformation [Eqn. (2)] and another is the geometric mean regression (GMR) (Clark and Perry 1995, Warton et al 2007). The advantage of the first approach is its computational simplicity and the advantage of the second or GMR is that it is more robust for small sample size (*N*<15 according to Clark and Perry 1995). Both approaches preserve the scale invariance of power law. Regarding the second aspect or the interpretation of the TPL, there is controversy on the claim that TPL parameter (*b*) is species specific, in particular when there are changes in sampling method, life stage, environment or spatial scale (Taylor *et al*. 1988, Clark and Perry 1995). Our opinion is that, unlike parameter *a*, parameter *b* is primarily shaped by evolutionary forces and less influenced by ecological or environmental factors. However, we do not take the "invariance" or "constancy" at ecological time scale or with environmental factors as granted. Instead, we draw conclusions based on rigorous statistical tests of the differences in parameter *b* among treatments. In the case of this study, we perform the *permutation* (*randomization*) test to judge whether or not the TPL parameters are invariant. For further information on the randomization test, readers are referred to Collingridge (2013).

To further harness the TPL parameters, Ma (1991, 2012a, 2015) derived a third parameter ($M_0$) for TPL or its extensions at the community scale, *population aggregation critical density* at the population scale or *community critical heterogeneity* at the community scale, which is in the form of:

$$M_0 = \exp[\ln(a)/(1-b)] \quad (a > 0, \ b \neq 1), \quad (3)$$

where *a* & *b* are TPL parameter. $M_0$ is the level of *mean* population abundance, the COVID-19 or SARS infection level in the case of this study, at which the fluctuation (dynamics) of virus infection is random (following Poisson statistical distribution generally). When $m>M_0$, the population (infection) fluctuation (dynamics) is more *irregular* than random (often following highly skewed distributions such as the negative binomial distribution or power-law statistical distribution). In this case, population is highly unstable and the infection may expand continuously. When $m<M_0$, the infection fluctuation is *regular* and may follow the *uniform* statistical distribution. In this case, the inflection level should be stabilized or might even die off. When $m=M_0$, the infection is random and should follow Poisson distribution statistically. In the context of this study, we term PACD or $M_0$ as *mean* infection critical threshold or mean infection threshold, which is similar to classic Allee effects (Allee 1927). Although the both are similar conceptually, it should be emphasized that they have different ecological interpretations. In particular, when $m=M_0$, it means that the population spatial distribution or temporal stability (variability) is random; when $m<M_0$, the distribution or stability is regular or uniform. Whether or not the population may die off is uncertain. If $m$ happens to coincide with the threshold of Allee effects, population (infections) may indeed die off.

**Approximating metapopulation with metacommunity model: Hubbell's (2001) UNTB and Harris *et al.* (2015) HDP-MSN model**

Hubbell's (2001) UNTB (Unified Neutral Theory of Biodiversity) conceptually distinguishes between local community dynamics and metacommunity dynamics, both of which are assumed to be driven by similar neutral processes—stochastic drifts in species demography, local speciation and global dispersal (migration). The UNTB has two key parameters (elements): (*i*) the immigration rate ($I_i$) that controls the coupling of a local community to the metacommunity; (*ii*) the speciation rate (also known as the fundamental biodiversity number $\theta$) that can be interpreted as the rate at which new individuals are added to the metacommunity due to speciation. The UNTB assumes that the SAD (species abundance distribution) of each community sample can be described by the multinomial (MN) distribution, which is parameterized by the previously mentioned two parameters. Testing the UNTB model is then computationally equivalent to testing the goodness-of-fitting to the MN distribution. However, a fully general case of fitting multiple sites UNTB with different immigration rates is computationally extremely challenging (actually intractable) even for small number of sites. In other words, to test the UNTB in truly multi-site (multiple local communities), approximate algorithms must be used (Harris *et al*. 2015). It was Harris *et al*. (2015) that developed an efficient Bayesian fitting framework for testing the UNTB by approximating the neutral models with the hierarchical Dirichlet process (HDP). For this reason, we refer to the computational framework developed by Harris *et al.* (2015) as HDP-MSN (hierarchical Dirichlet process—multi-site neutral) model.

It was found that for large local population sizes, assuming a fixed finite-dimensional metacommunity distribution with *S* species present, the local community distribution ($\pi_i$) could be approximated by a *Dirichlet* distribution (Sloan et al 2006 & 2007). Based on this finding, Harris *et al*. (2015) developed their computationally efficient, general framework for approximating the UNTB. Assuming a potentially infinite number of species can be observed in the local community, the stationary distribution of observing local population *i* can be modeled with a Dirichlet process (DP), *i.e.*,

$$\bar{\pi}_i | I_i, \bar{\beta} \sim DP(I_i, \bar{\beta}) \tag{4}$$

where $\bar{\beta} = (\beta_1, ..., \beta_S)$ is the relative frequency of each species in the metacommunity.

At the metacommunity level, a Dirichlet process is also applicable and the metacommunity distribution can be modeled with a stick breaking process, *i.e.*,

$$\bar{\beta} \sim Stick(\theta) \tag{5}$$

Given that both local community and metacommunity follow Dirichlet processes, the problem can be formulated as a hierarchical Dirichlet process (HDP) in the domain of machine learning (Teh *et al.* 2006, Harris *et al.* 2015).

Furthermore, Dirichlet process (DP) can be formulated as the so-called Chinese restaurant process, from which Antoniak equation (Antoniak 1974) can be derived. The Antoniak equation represents for the number of types (or species) (*S*) observed following *N* draws from a Dirichlet process with concentration parameter $\theta$, and is with the following form:

$$P(S|\theta, N) = s(N, S) \theta^S \frac{\Gamma(\theta)}{\Gamma(\theta + N)} \tag{6}$$

where $s(N, S)$ is the unsigned Stirling number of the first kind and $\Gamma(.)$ is the gamma function.

By combining previous equations (4-6) and the previously mentioned multi-nominal (MN) distribution of the community samples, Harris *et al.* (2015) obtained their full HDP-MSN model (hierarchical Dirichlet process—multisite neutral). They further developed an efficient Gibbs sampler for the UNTB-HDP approximation, which is a type of Bayesian Markov Chain Monte Carlo (MCMC) algorithm.

By treating the coronavirus infections at different sites (*e.g.*, provinces of China, or different countries/regions of the world, in this study) as a "metacommunity" consisting of *N* local communities (*e.g.*, each local community corresponding to a province), the above-described metacommunity model can be built with the dataset of daily incremental infections, nationally or internationally. Different from traditional metacommunity concept, here the "metacommunity" is actually a metapopulation consisting of *N* local populations. However, if we treat the local infections at a particular time point (*e.g.*, day) as a local sub-population, then the virus sub-populations at different time points can be considered as a total population (or "species" in the terminology of community ecology). With this conceptual transformation, the concept and models for metacommunity and Hubbell's UNTB can be readily applied to the metapopulation of coronavirus infections without a need to revise the models. With this adaptive scheme, the fundamental biodiversity number (speciation rate: $\theta$) from the previously introduced HDP-MSN model can be used to approximate the average local (contagion) infection rate. Similarly, the fundamental dispersal number ($M$) can be used to approximate the average infection rate through migration. The migration probability ($m$), which is a function of $M$, has a similar interpretation as $M$, but simply in the form of probability.

As a side note, we expect that the coronavirus infections should follow Hubbell's (2001) UNTB theory, which is not surprising to us, due to our treatment of metapopulation as metacommunity. This is because the "species" in our "metacommunity" are, in fact, populations of a single virus species, and they should be "equivalent" in terms of the neutral theory. Therefore, whether or not the coronavirus infections satisfy the neutral theory model, specifically Harris *et al*. (2015) HDP-MSN model, is not our focus. Instead, we focus on the estimated parameters from HDP-MSN modeling and their interpretations to explore their implications to the outbreak of virus infections. Of course, if the neutral theory model fails to fit the virus infections data statistically, we stop pursuing the interpretations of their parameters.

## Results

From Tables 1-2, which were summarized from Tables S1-S8 in the OSI (online supplementary information), we obtain the following findings, which are also illustrated with Figs 1-7.

(*i*) The spatiotemporal fluctuation scaling (changes) of COVID-19 infections follow TPL (Taylor's power law) at all scales (schemes) tested, including world-wide, country-wide, cumulative and daily incremental infections, as well as gradually shrunken *partial* datasets to test TPL robustness, as evidenced by the $p$-value<0.001 from the TPL fittings. The brief results were summarized in Table 1 and the detailed results were displayed in Tables S1-S2 and S4-S7. It was discovered that the infections of COVID-19, like populations of other organisms, follow seeming universal power law. This implies that the infection of the novel coronavirus is highly contagious and its outbreak (spread) is chaotically unstable in general, as indicated by the $b$-values exceeding *1* (for daily incremental infection) or exceeding *2* (for cumulative infections).

(*ii*) The TPL aggregation (stability) parameter (*b*) for COVID-19 seems rather stable or even invariant, as evidenced by the randomization tests (Tables S3 & S8). The randomization tests were performed by statistically comparing the model parameters fitted with the whole datasets and those fitted with the "partial datasets." The schemes of partial datasets were devised by removing 1, 2, … ... 15 days data from either head or tail of the dataset to *test* the robustness of TPL model and to *test* the invariance of the model parameters. It was found that, while TPL-*b* seems rather stable and even invariant with disease kinds (COVID-19 or SARS). Instead, the TPL parameters *a* & $M_0$ seem variable, suggesting their potential roles in assessing and interpreting the stability of the virus infections. For example, the difference in TPL-*b* values for worldwide COVID-19 and SARS infections (2.076 *vs.* 2.023) is not statistically significant (*p*-value=0.874 (Table S8 for cumulative TPL modeling). Therefore, we postulate that the TPL-*b* may be coronavirus specific and may be primarily shaped by evolutionary forces.

(*iii*) The population aggregation critical density (PACD or $M_0$) (*i.e.*, infection critical threshold) seems to depend on disease kinds (COVID-19 or SARS), space (regions), and time (stage of disease outbreak), as indicated by the randomization test results (Table S3 & S8). Given that $M_0$ measures the infection threshold at which infections are random (or follow Poisson distribution), a lower $M_0$ may indicate a lower infection "tolerance" level since crossing the level of $M_0$ may signal the highly variable (unstable) infections. For example (see Table 1), $M_0$=5.758 for COVID-19 in China, $M_0$=2.701 for COVID-19 worldwide, $M_0$=9.475 for SARS worldwide may suggest that the infection tolerance level of COVID-19 in China is about twice higher than that of the worldwide, but about ½ lower than that of worldwide SARS infections. In other words, the COVID-19 infections seem more "dangerous" than SARS from a public health perspective.

Note that $M_0$ is the *mean* infection level; therefore, it can be more meaningful to convert it into absolute value from a biomedical perspective. For example, $M_0$=5.758 for COVID-19 in China, when converted to total national infection level, would be 5.758x34 (provinces)=196. This threshold number may suggest that when the number of total infections nationally is at this level, the infections are random. When infections exceed this threshold level, the infections can be non-random and highly unstable. When infections are below this threshold level, the infection should be rather stable (following a Uniform distribution statistically) or may even die off if the threshold level coincides with the level of Allee effects. Since the level of Allee effects is still unknown, whether or not the infections under $M_0$ will die off is still an open question.

(***iv***) As expected, all datasets passed the neutrality tests of HDP-MSN model as indicated by the $p$-value>0.05 (Table 2). We use the ratio of $Q=M/\theta$ as a measure of the relative importance of "infection spread *via* migration" *vs*. "infections spread *via* local contagion" in spreading the infections, with larger $Q$ indicating higher migration role and smaller $Q$ indicating higher local contagion. The ratio of $Q=133.581/6.325≈21$ (Table 2) indicating that spread *via* migration is approximately 21 times more significant than spread *via* local contagion on average nationally in China. However, the worldwide $Q$ ratio is approximately $Q=1$ (8.037/10.801). Therefore, the ratio $Q$ is dependent on time (disease stages), space (regions), and disease-kinds (COVID-19 or SARS), and perhaps most importantly, public-health interventions such as quarantines and/or mobility control.

Compared with COVID-19, SARS appeared to exhibit a different pattern of the relative importance of migration *vs*. local contagion in spreading the infections. We postulate that this difference might signal the higher risk of pandemics of COVID-19 compared with SARS. Nevertheless, we cannot exclude the possibility that the range of SARS datasets were complete, while COVID-19 infections are not over yet.

The third parameter ($m$) or immigration probability suggests the level of infection *via* migration. The $m=0.052$ in China *vs*. $m=0.003$ worldwide for COVID-19 suggested that the risk of infection *via* migration within China is approximately 10 times higher than that of worldwide migration. This is most likely due to the disruption of international travels. Comparing the $m$ for COVID-19 and SARS for worldwide data (0.003 for COVID-19 *vs*. 0.040 for SARS) may simply be due to stronger travel restrictions imposed for controlling COVID-19 outbreaks. Similar to the above comparison, this difference may be due to the difference in the data range.

(***v***) Comparing the PACD ($M_0$) (Table 1) and $\theta$ in (Table 2) exhibited a very interesting phenomenon. For the cumulative infections in China, both parameters ($M_0$ and $\theta$) are rather close to each other. For example, the difference between $M_0=5.758$ of COVID-19 in China (Table 1) and $\theta=6.325$ of COVID-19 in China (Table 2) is only approximately 0.6. We postulate that, when the "local speciation (contagion)" ($\theta$) approximates the population aggregation critical density or infection critical threshold ($M_0$), the infections could become random (suggesting a potentially stabilized infection level). That is, without inputs from external migration, local infections *via* local contagion (measured by $\theta$) could become random (as indicated by $M_0$). If this postulation is true, then it may suggest that the mobility control (such as travel restrictions or quarantines) can be critically effective in stabilizing outbreaks. This finding indicates the complementary nature of the two approaches we adopted in this study. Nevertheless, it is important to reiterate that this closeness between $M_0$ and $\theta$ ($M_0 \approx \theta$) is likely to be an exception, rather than the "norm", for the reasons explained below.

As to the lack of closeness between $M_0$ and $\theta$ of COVID-19 at worldwide scale ($M_0$=2.701 *vs*. $\theta$=10.801), this may indicate that control mobility is not sufficient to stabilize infections world widely anymore at the current stage. In the case of SARS, the worldwide $M_0$=9.475 *vs*. $\theta$=27.169, indicated that local contagion ($\theta$) alone already exceed the infection critical threshold ($M_0$). Therefore, $\theta >> M_0$ (PACD) may be the "norm."

## Conclusions and Discussion

We aimed to discover critical insights on the endemic/epidemic/pandemic characteristics of the coronavirus outbreaks. Methodically, we first apply the TPL (Taylor 1961, 2019) scaling law for measuring spatial (temporal) heterogeneity (stability) to shed important lights on the *stability* of coronavirus infection outbreak. We then apply Hubbell's (2001) UNTB (specifically Harris *et al*. 2015 HDP-MSN model) to approximate the metapopulation of coronavirus outbreak and to estimate the critical metapopulation parameters that may offer insights on the relative importance of *migration* vs. local *contagion* in driving or suppressing virus outbreaks. Furthermore, both TPL and metapopulation modeling complemented each other and revealed the five findings explained in the previous sections, from which we further summarize the following insights and discuss their implications below.

First, the coronavirus infections, including both COVID-19 and SARS appear to possess characteristic *stability* parameter (*b*) values, slightly exceeding *2* in terms of the daily cumulative infections and exceeding *1* in terms of the daily incremental infections. As suggested by the literature of TPL, when the TPL parameter (*b*) exceeds *2*, the population dynamics can be rather chaotic, indicating potentially chaotic behavior of coronavirus outbreaks. The TPL-*b* values for both the coronaviruses are at the high end of the *b*-value range in population ecology, compared with the *b*-values of most microbial and macrobial species in existing literature, which are usually *b*<2.

While TPL-*b* appears rather stable or even invariant with time and/or space, the population aggregation critical density (PACD) (*i.e.*, $M_0$ or infection critical threshold), which is the level at which infections are random and below which infections may be stabilized, can depend on disease kinds (COVID-19 or SARS), time (disease or outbreak stages), and space (regions). We postulate that $M_0$ should also be influenced by public-health interventions such as quarantines and travel restrictions. Therefore, $M_0$ can be an important epidemiological parameter for evaluating the characteristics of disease outbreaks. For example, COVID-19 exhibited significantly lower threshold ($M_0$) than SARS, suggesting a potentially lower infection "tolerance" threshold of COVID-19. Nevertheless, it should be cautioned that the "tolerance" threshold only means the level of random infections, which may signal the level of stabilized infections. However, whether or not the "tolerance" threshold is biometrically tolerable or safe depends on other biomedical characteristics, among which Allee effects can be a critical factor to determine whether or not the infections will die off or persist. We postulate that if $M_0$ coincide or is below the level Allee effects in action, the infections may die off.

Second, all datasets we tested easily passed the neutrality test with HDP-MSN and indicated that the approximation of the metapopulation with metacommunity in the case of coronavirus infections is feasible. Two parameters from the HDP-MSN, *i.e.*, fundamental biodiversity number ($\theta$) and fundamental dispersal number ($M$) measure the average infections from local "speciation" (*i.e.*, *local contagion*) and average infections from "dispersal" (migration) from other local populations. In other words, $\theta$ measures the spread level from local contagion and $M$ measures spread level from external migration. A third parameter migration probability ($m$) is a function of fundamental dispersal number ($M$), measures the probability of infections from migration. All three parameters can depend on disease kinds (COVID-19 or SARS), time (disease stage), space (regions), as well as public-health interventions (such as quarantines, mobility control and/or isolation for treatments).

The ratio of $Q=M/\theta$ may be used as a measure of the relative importance of "infection spread *via* migration" *vs*. "infections spread *via* local contagion" in spreading the infections, with larger $Q$ indicating higher migration role and smaller $Q$ indicating higher local contagion.

Third, both the TPL scaling law and metapopulation modeling may complement each other. The difference (closeness) between $M_0$ (the infection critical threshold from TPL) and $\theta$ (local *contagion* or "speciation" rate from metapopulation modeling) may signal the effectiveness of completely blocking the migration (dispersal) in spreading infections. For example, in the case of COVID-19 infections in China, both parameters ($M_0$ and $\theta$) are rather close to each other and their difference is only approximately 0.6, suggesting that, without external inputs, the infections from local contagion is only approximately 0.6 higher than the infection critical threshold ($M_0$). This makes the mission of controlling local contagion for *stabilizing* infections much less challenging than the mission when $\theta \gg M_0$.

Finally, we suggest that the approaches demonstrated previously should be of general applicability for epidemiological research. In particular, we consider TPL-*b* can be a pathogen specific parameter, primarily shaped by evolutionary forces. Another TPL parameter, Ma (1991, 2015) PACD (population aggregation critical density) or infection critical threshold ($M_0$), as renamed in this study in the context of coronavirus infections, may be particular useful for assessing critical threshold for stabilizing infections (outbreaks). A lower $M_0$ may signal a lower threshold for the occurrence of outbreak, which may imply higher outbreak risk from a public health perspective. Previously, we have shown that COVID-19 appears to have a higher outbreak risk compared with SARS. Given that both of the coronaviruses appear to have virtually equal *b*-values, then one wonders if they have any essential differences, revealed in this study, from the epidemiological or public-health perspectives. $M_0$ (infection critical threshold) seems to be such an effective indicator (property) for assessing outbreak threshold (risk). Our study suggested that COVID-19 possesses a lower infection critical threshold ($M_0$) than SARS does may be explained by its lower fatality rate compared with SARS. Obviously lower lethality to hosts should actually be favorable for the survival and spread of virus.

## Acknowledgements and Disclaims
I acknowledge the data collection and computational support from Lianwei Li and Wendy Li from the Computational Biology and Medical Ecology Lab, the Chinese Academy of Sciences. I am responsible for all the interpretations of the results, including possible errors. However, I am not responsible for any direct or indirect inferences from this article, including those from partial quotes of the text, figures and/or tables.

## Data Availability
The COVID-19 (confirmed) infection datasets during January 19 and February 29 were collected from (https://news.qq.com/zt2020/page/feiyan.htm#/) and (https://news.ifeng.com/c/special/7tPlDSzDgVk). The worldwide SARS infections were from the WHO (https://www.who.int/csr/sars/country/en/).

# Tables 1-2

**Table 1**. The TPL (Taylor's power law) parameters fitted to the datasets of COVID-19 infections and SARS infections, respectively, with log-transformed linear regression (LLR) and geometric mean regression (GMR), respectively*

| Datasets | | Log-Liner Regression (LLR) | | | | | Geometric Mean Regression (GMR) | | | N |
|---|---|---|---|---|---|---|---|---|---|---|
| | | $b$ | $\ln(a)$ | $M_0$ | $R$ | $P$-value | $b$ | $\ln(a)$ | $M_0$ | |
| **COVID-19** | | | | | | | | | | |
| Cumulative | China | 2.153 | -2.018 | 5.758 | 0.995 | 0.000 | 2.163 | -2.070 | 5.930 | 33 |
| | China except for Hubei | 2.108 | -1.806 | 5.103 | 0.993 | 0.000 | 2.123 | -1.879 | 5.333 | 32 |
| | Hubei Province | 2.292 | -2.990 | 10.109 | 0.996 | 0.000 | 2.302 | -3.050 | 10.441 | 17 |
| | World | 2.076 | -1.069 | 2.701 | 0.944 | 0.000 | 2.198 | -1.381 | 3.167 | 37 |
| | World except for China | 2.069 | -1.055 | 2.683 | 0.910 | 0.000 | 2.274 | -1.532 | 3.330 | 36 |
| Daily Incremental | China | 1.902 | 0.373 | 0.661 | 0.985 | 0.000 | 1.932 | 0.315 | 0.713 | 33 |
| | China except for Hubei | 1.753 | 0.596 | 0.453 | 0.980 | 0.000 | 1.789 | 0.532 | 0.509 | 32 |
| | Hubei Province | 1.912 | 0.783 | 0.424 | 0.979 | 0.000 | 1.952 | 0.653 | 0.504 | 17 |
| | World | 1.820 | 1.598 | 0.142 | 0.980 | 0.000 | 1.857 | 1.576 | 0.159 | 36 |
| | World except for China | 1.797 | 1.596 | 0.135 | 0.972 | 0.000 | 1.848 | 1.575 | 0.156 | 35 |
| **SARS** | | | | | | | | | | |
| Cumulative | World | 2.020 | -2.293 | 9.475 | 0.965 | 0.000 | 2.094 | -2.452 | 9.408 | 23 |
| | World except for China | 1.902 | -2.112 | 10.409 | 0.931 | 0.000 | 2.042 | -2.372 | 9.744 | 22 |
| Daily Incremental | World | 1.546 | 2.829 | 0.006 | 0.906 | 0.000 | 1.707 | 3.148 | 0.012 | 18 |
| | World except for China | 1.537 | 2.806 | 0.005 | 0.815 | 0.000 | 1.887 | 3.630 | 0.017 | 17 |

*Given the minor differences from both LLR and GMR approaches and $N>15$, we adopted the results from LLR.
**$P$-value<0.001 indicated that all the model fittings are statistically significant. $R$ (linear correlation coefficient) further confirmed exceptional significance level of TPL fittings.

**Table 2.** The meta-population model parameters fitted to the daily increments of COVID-19 and SARS infections, respectively, by adapting Hubbell's UNTB (unified neutral theory of biodiversity) model*

| Datasets | $\theta$ | $M$ | $m$ | Metapopulation | | | Local Population | | |
|---|---|---|---|---|---|---|---|---|---|
| | | | | $N_M$ | $N$ | $P_M$ | $N_L$ | $N$ | $P_L$ |
| **COVID-19** | | | | | | | | | |
| China | 6.325 | 133.581 | 0.052 | 2452 | 2500 | 0.981 | 1988 | 2500 | 0.795 |
| China except for Hubei | 5.879 | 139.983 | 0.257 | 2467 | 2500 | 0.987 | 1005 | 2500 | 0.402 |
| Hubei Province | 7.076 | 79.451 | 0.020 | 2394 | 2500 | 0.958 | 1904 | 2500 | 0.762 |
| World | 10.801 | 8.037 | 0.003 | 2427 | 2491 | 0.974 | 1736 | 2491 | 0.697 |
| World except for China | 10.330 | 7.857 | 0.035 | 2316 | 2492 | 0.929 | 1501 | 2491 | 0.603 |
| **SARS** | | | | | | | | | |
| World | 27.169 | 16.241 | 0.040 | 2439 | 2494 | 0.978 | 2148 | 2494 | 0.861 |
| World-except China | 32.219 | 15.094 | 0.274 | 2363 | 2495 | 0.947 | 2144 | 2495 | 0.859 |

*$\theta$: Local contagion rate; $M$: Mean migration rate; $m$: immigration probability; $N=2500$ is the number of Gibb samples selected from 25000 simulated communities performed to test the fitting of Harris *et al*. (2015) HDP-MSN model; $N_M$ & $N_L$ are the number of simulations that passed the neutrality test (by comparing with actual daily incremental infection numbers), respectively. $P_M$ and $P_L$ are the *pseudo-P* values from performing the neutrality test at metapopulation and local population levels, respectively. When *pseudo-P* value>0.05, it indicates that the dataset satisfied the neutral model, and the parameters ($\theta$, $M$, and $m$) can be harnessed to assess and interpret the relative importance of local contagion vs. migration in driving outbreaks of coronaviruses infections.

**Figures 1-2**

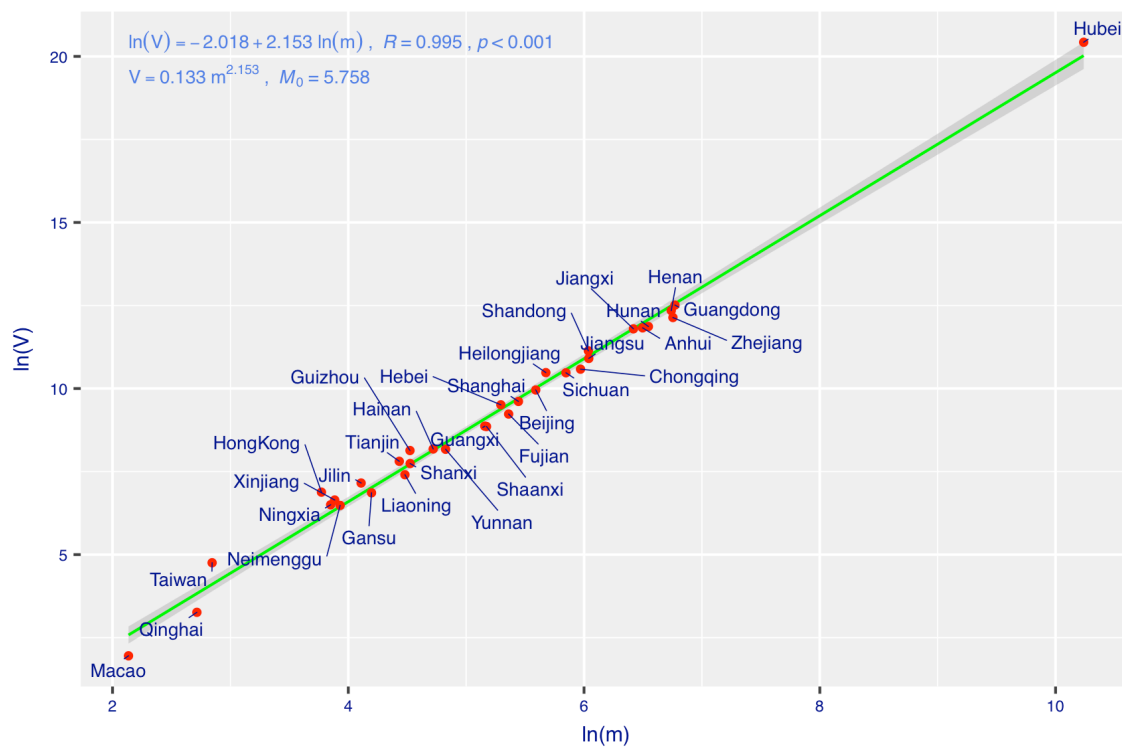

**Fig 1.** TPL (Taylor's power law) model fitted to the cumulative infections of COVID-19 in China (Infection Critical Threshold $M_0$=5.758; Spatiotemporal Scaling Parameter $b$=2.153; $P$-value<0.001)

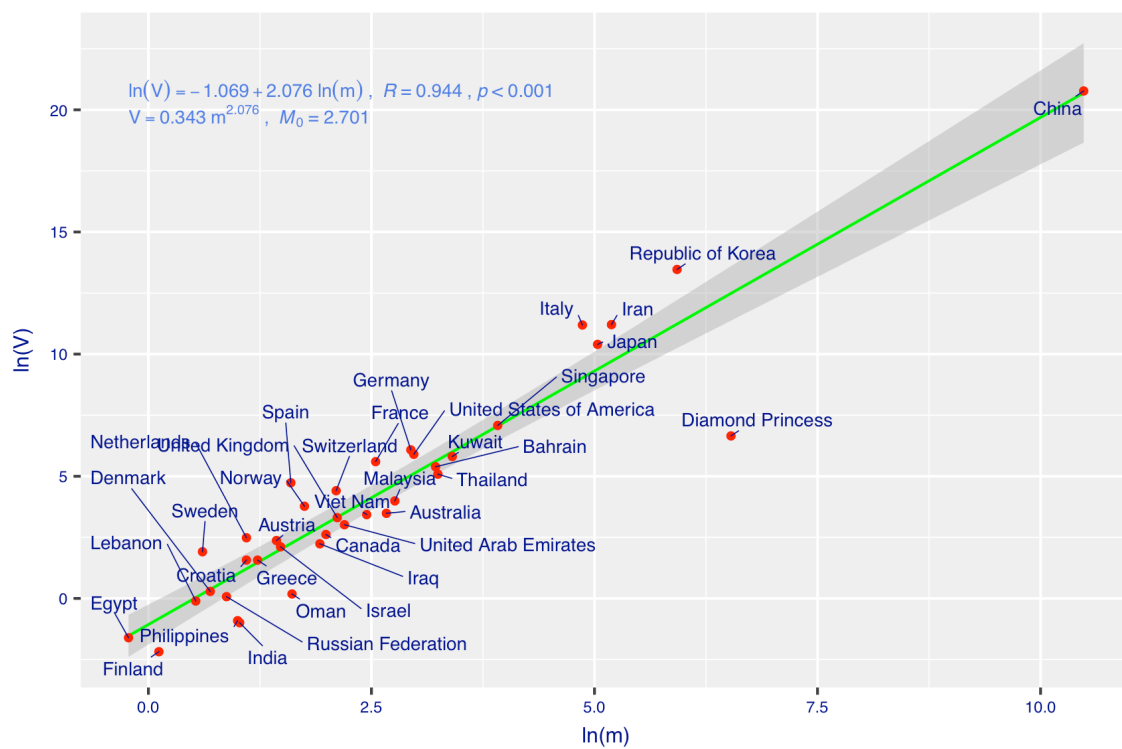

**Fig 2.** TPL (Taylor's power law) model fitted to the cumulative infections of COVID-19 worldwide (Infection Critical Threshold $M_0$=2.701; Spatiotemporal Scaling Parameter $b$=2.076; $P$-value<0.001)

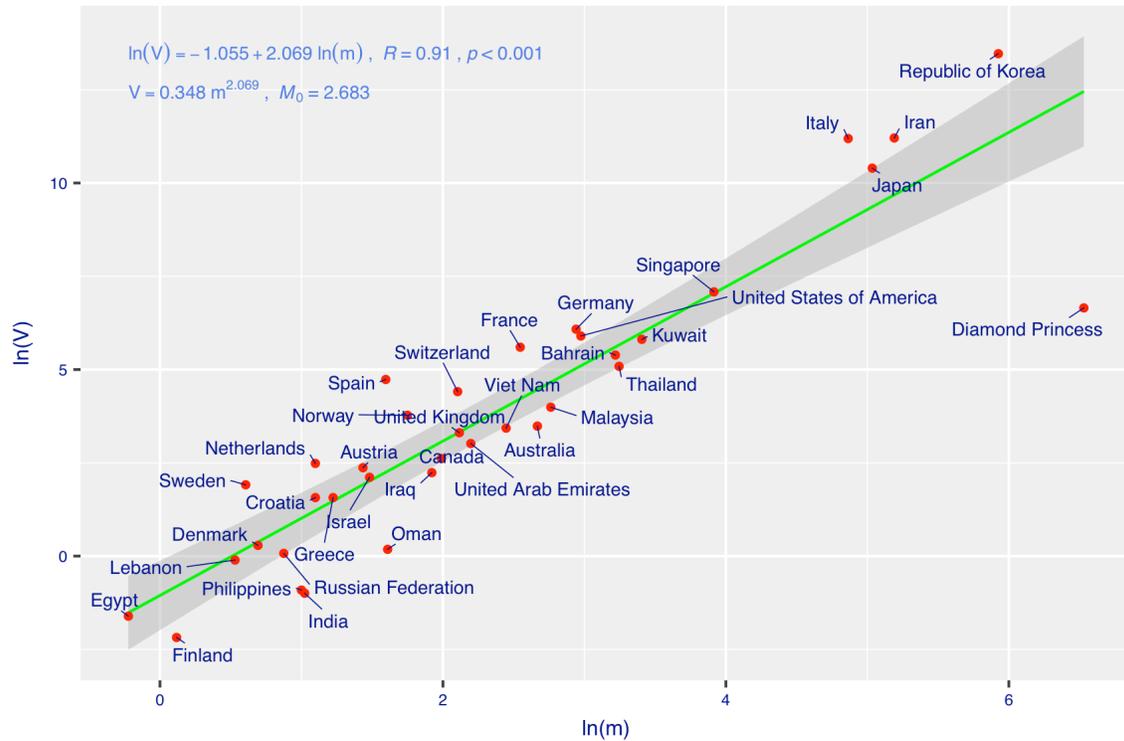

**Fig 3.** TPL (Taylor's power law) model fitted to the cumulative infections of COVID-19 worldwide except for China (Infection Critical Threshold $M_0$=2.683; Spatiotemporal Scaling Parameter $b$=2.069; $P$-value<0.001)

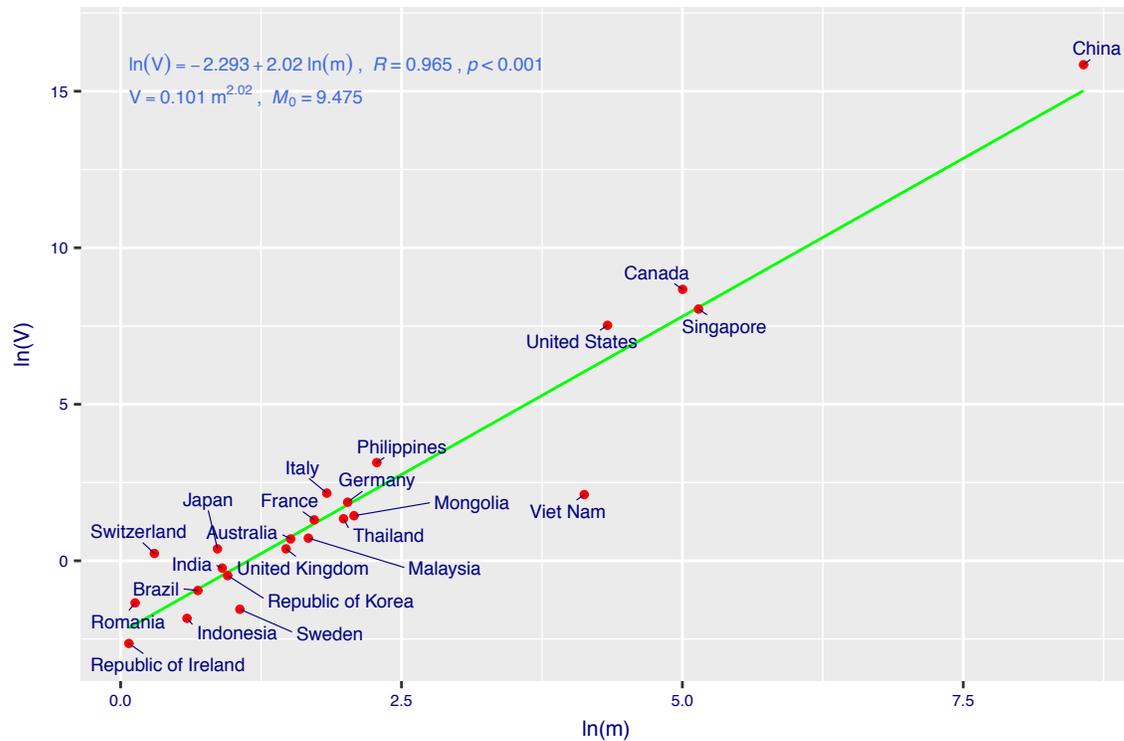

**Fig 4.** TPL (Taylor's power law) model fitted to the cumulative infections of 2003-SARS worldwide (Infection Critical Threshold $M_0$=9.475; Spatiotemporal Scaling Parameter $b$=2.020; $P$-value<0.001)

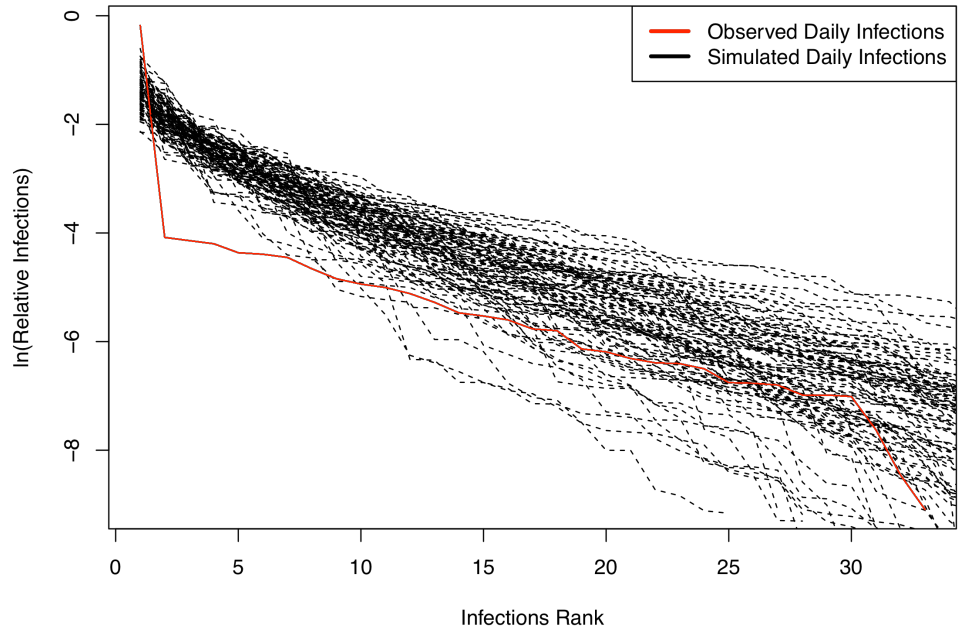

**Fig 5**. Approximating the metapopulation of daily incremental COVID-19 infections in China with Hubbell's (2001) UNTB, specifically Harris *et al*. (2015) HDP-MSN model.

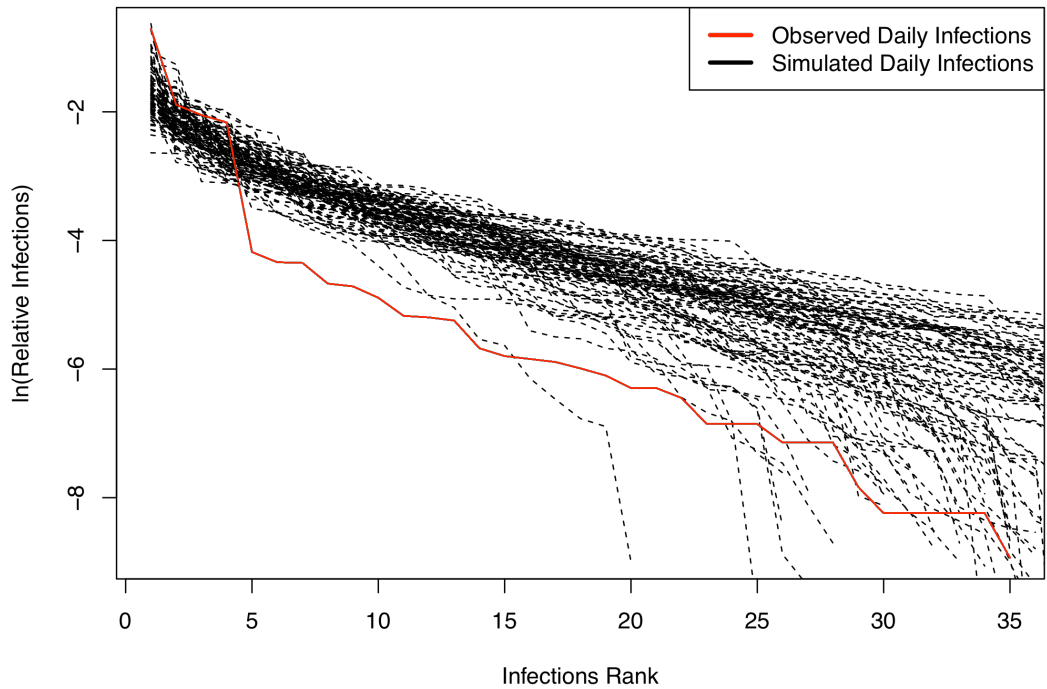

**Fig 6**. Approximating the metapopulation of daily incremental COVID-19 infections worldwide except for China with Hubbell's (2001) UNTB, specifically Harris *et al*. (2015) HDP-MSN model.

# List of Online Supplementary Tables 1-8

**Table S1A.** Fitting the TPL to the daily *cumulative* COVID-19 infections in the whole China with two schemes: fitting with the full datasets or fitting with the gradually shrinking datasets

**Table S1B.** Fitting the TPL to the daily *cumulative* COVID-19 infections in the whole China except for Hubei province with two schemes: fitting with the full datasets or fitting with the gradually shrinking datasets

**Table S1C**. Fitting the TPL to the daily *cumulative* COVID-19 infections in the Hubei province with two schemes: fitting with the full datasets or fitting with the gradually shrinking datasets

**Table S2A.** Fitting the TPL to the daily *incremental* COVID-19 infections in the whole China with two schemes: fitting with the full datasets or fitting with the gradually shrinking datasets

**Table S2B.** Fitting the TPL to the daily *incremental* COVID-19 infections in the whole China except for Hubei province with two schemes: fitting with the full datasets or fitting with the gradually shrinking datasets

**Table S2C.** Fitting the TPL to the daily *incremental* COVID-19 infections in Hubei province with two schemes: fitting with the full datasets or fitting with the gradually shrinking datasets

**Table S3.** The randomization (permutation) test results of the TPL stability model parameters

**Table S4A.** The results of fitting TPL to the *cumulative* infections of COVID-19 worldwide with two schemes: fitting with the full datasets or fitting with the gradually shrinking datasets

**Table S4B.** The results of fitting TPL to the *cumulative* infections of COVID-19 worldwide except for China with two schemes: fitting with the full datasets or fitting with the gradually shrinking datasets

**Table S5A.** The results of fitting TPL to the *daily incremental* infections of COVID-19 worldwide with two schemes: fitting with the full datasets or fitting with the gradually shrinking datasets

**Table S5B.** The results of fitting TPL to the *daily incremental* infections of COVID-19 worldwide except for China with two schemes: fitting with the full datasets or fitting with the gradually shrinking datasets

**Table S6A**. Fitting the TPL to the *cumulative SARS* infection worldwide with two schemes: fitting with the full datasets or fitting with the gradually shrinking datasets

**Table S6B**. Fitting the TPL to the *cumulative SARS* infection worldwide except for China with two schemes: fitting with the full datasets or fitting with the gradually shrinking datasets

**Table S7A.** Fitting the TPL to the *daily incremental SARS* infection worldwide with two schemes: fitting with the full datasets or fitting with the gradually shrinking datasets

**Table S7B.** Fitting the TPL to the *daily incremental SARS* infection worldwide except for China with two schemes: fitting with the full datasets or fitting with the gradually shrinking datasets

**Table S8.** The randomization (permutation) test for the TPL stability model parameters

Online Supplementary Tables for "**Spatiotemporal fluctuation scaling law and metapopulation modeling of the novel coronavirus (COVID-19) outbreak**"

**Table S1A.** Fitting the TPL to the daily *cumulative* COVID-19 infections in the whole China with two schemes: fitting with the full datasets or fitting with the gradually shrinking datasets

| Treatments | $b$ | $\ln(a)$ | $M_0$ | $R$ | $p$-value | $N_{area}$ | $T_{day}$ |
|---|---|---|---|---|---|---|---|
| **Temporal Stability Model** | | | | | | | |
| Full Data between Jan 19 and Feb 29 | 2.153 | -2.018 | 5.758 | 0.995 | 0.000 | 33 | 42 |
| Minus tail 1 | 2.169 | -2.229 | 6.731 | 0.994 | 0.000 | 33 | 41 |
| Minus tail 2 | 2.182 | -2.437 | 7.852 | 0.993 | 0.000 | 33 | 40 |
| Minus tail 3 | 2.198 | -2.670 | 9.280 | 0.992 | 0.000 | 33 | 39 |
| Minus tail 4 | 2.224 | -2.973 | 11.343 | 0.990 | 0.000 | 33 | 38 |
| Minus tail 5 | 2.242 | -3.240 | 13.596 | 0.987 | 0.000 | 33 | 37 |
| Minus tail 6 | 2.248 | -3.456 | 15.926 | 0.986 | 0.000 | 33 | 36 |
| Minus tail 7 | 2.259 | -3.708 | 19.029 | 0.983 | 0.000 | 33 | 35 |
| Minus tail 8 | 2.292 | -4.115 | 24.165 | 0.977 | 0.000 | 33 | 34 |
| Minus tail 9 | 2.330 | -4.571 | 31.067 | 0.969 | 0.000 | 33 | 33 |
| Minus tail 10 | 2.409 | -5.291 | 42.703 | 0.952 | 0.000 | 33 | 32 |
| Minus tail 11 | 2.244 | -4.491 | 36.909 | 0.971 | 0.000 | 32 | 31 |
| Minus tail 12 | 2.264 | -4.874 | 47.245 | 0.965 | 0.000 | 32 | 30 |
| Minus tail 13 | 2.137 | -4.364 | 46.451 | 0.959 | 0.000 | 31 | 29 |
| Minus tail 14 | 2.142 | -4.690 | 60.661 | 0.952 | 0.000 | 31 | 28 |
| Minus tail 15 | 2.148 | -5.035 | 80.324 | 0.944 | 0.000 | 31 | 27 |
| | | | | | | | |
| Minus head 1 | 2.155 | -1.999 | 5.647 | 0.995 | 0.000 | 33 | 41 |
| Minus head 2 | 2.157 | -1.982 | 5.542 | 0.996 | 0.000 | 33 | 40 |
| Minus head 3 | 2.159 | -1.960 | 5.422 | 0.996 | 0.000 | 33 | 39 |
| Minus head 4 | 2.161 | -1.936 | 5.299 | 0.996 | 0.000 | 33 | 38 |
| Minus head 5 | 2.163 | -1.914 | 5.184 | 0.996 | 0.000 | 33 | 37 |
| Minus head 6 | 2.165 | -1.891 | 5.069 | 0.996 | 0.000 | 33 | 36 |
| Minus head 7 | 2.168 | -1.868 | 4.953 | 0.996 | 0.000 | 33 | 35 |
| Minus head 8 | 2.170 | -1.842 | 4.830 | 0.996 | 0.000 | 33 | 34 |
| Minus head 9 | 2.171 | -1.812 | 4.699 | 0.996 | 0.000 | 33 | 33 |
| Minus head 10 | 2.173 | -1.783 | 4.576 | 0.996 | 0.000 | 33 | 32 |
| Minus head 11 | 2.175 | -1.754 | 4.451 | 0.996 | 0.000 | 33 | 31 |
| Minus head 12 | 2.177 | -1.723 | 4.323 | 0.996 | 0.000 | 33 | 30 |
| Minus head 13 | 2.179 | -1.690 | 4.196 | 0.996 | 0.000 | 33 | 29 |
| Minus head 14 | 2.180 | -1.654 | 4.062 | 0.996 | 0.000 | 33 | 28 |
| Minus head 15 | 2.181 | -1.618 | 3.935 | 0.996 | 0.000 | 33 | 27 |

**Table S1B.** Fitting the TPL to the daily *cumulative* COVID-19 infections in the whole China except for Hubei province with two schemes: fitting with the full datasets or fitting with the gradually shrinking datasets

| Treatments | $b$ | $\ln(a)$ | $M_0$ | $R$ | $p$-value | $N_{area}$ | $T_{day}$ |
|---|---|---|---|---|---|---|---|
| **Temporal Stability Model** | | | | | | | |
| Full Data between Jan 19 and Feb 29 | 2.108 | -1.806 | 5.103 | 0.993 | 0.000 | 32 | 41 |
| Minus tail 1 | 2.124 | -2.014 | 6.005 | 0.991 | 0.000 | 32 | 40 |
| Minus tail 2 | 2.135 | -2.211 | 7.014 | 0.990 | 0.000 | 32 | 39 |
| Minus tail 3 | 2.149 | -2.434 | 8.315 | 0.988 | 0.000 | 32 | 38 |
| Minus tail 4 | 2.177 | -2.746 | 10.305 | 0.984 | 0.000 | 32 | 37 |
| Minus tail 5 | 2.191 | -2.996 | 12.368 | 0.981 | 0.000 | 32 | 36 |
| Minus tail 6 | 2.189 | -3.166 | 14.335 | 0.978 | 0.000 | 32 | 35 |
| Minus tail 7 | 2.190 | -3.374 | 17.016 | 0.974 | 0.000 | 32 | 34 |
| Minus tail 8 | 2.225 | -3.783 | 21.968 | 0.964 | 0.000 | 32 | 33 |
| Minus tail 9 | 2.265 | -4.248 | 28.745 | 0.952 | 0.000 | 32 | 32 |
| Minus tail 10 | 2.364 | -5.065 | 41.027 | 0.926 | 0.000 | 32 | 31 |
| Minus tail 11 | 2.091 | -3.710 | 30.016 | 0.953 | 0.000 | 31 | 30 |
| Minus tail 12 | 2.099 | -4.033 | 39.240 | 0.944 | 0.000 | 31 | 29 |
| Minus tail 13 | 1.811 | -2.650 | 26.239 | 0.936 | 0.000 | 30 | 28 |
| Minus tail 14 | 1.789 | -2.823 | 35.844 | 0.924 | 0.000 | 30 | 27 |
| Minus tail 15 | 1.764 | -3.002 | 50.893 | 0.908 | 0.000 | 30 | 26 |
| | | | | | | | |
| Minus head 1 | 2.110 | -1.788 | 5.007 | 0.993 | 0.000 | 32 | 40 |
| Minus head 2 | 2.113 | -1.773 | 4.918 | 0.993 | 0.000 | 32 | 39 |
| Minus head 3 | 2.115 | -1.751 | 4.810 | 0.994 | 0.000 | 32 | 38 |
| Minus head 4 | 2.116 | -1.727 | 4.699 | 0.994 | 0.000 | 32 | 37 |
| Minus head 5 | 2.118 | -1.706 | 4.597 | 0.994 | 0.000 | 32 | 36 |
| Minus head 6 | 2.120 | -1.683 | 4.493 | 0.994 | 0.000 | 32 | 35 |
| Minus head 7 | 2.122 | -1.660 | 4.389 | 0.994 | 0.000 | 32 | 34 |
| Minus head 8 | 2.124 | -1.633 | 4.276 | 0.994 | 0.000 | 32 | 33 |
| Minus head 9 | 2.125 | -1.601 | 4.152 | 0.994 | 0.000 | 32 | 32 |
| Minus head 10 | 2.126 | -1.571 | 4.037 | 0.994 | 0.000 | 32 | 31 |
| Minus head 11 | 2.128 | -1.542 | 3.926 | 0.994 | 0.000 | 32 | 30 |
| Minus head 12 | 2.130 | -1.513 | 3.815 | 0.994 | 0.000 | 32 | 29 |
| Minus head 13 | 2.132 | -1.482 | 3.704 | 0.994 | 0.000 | 32 | 28 |
| Minus head 14 | 2.133 | -1.447 | 3.588 | 0.994 | 0.000 | 32 | 27 |
| Minus head 15 | 2.135 | -1.416 | 3.483 | 0.995 | 0.000 | 32 | 26 |

**Table S1C**. Fitting the TPL to the daily *cumulative* COVID-19 infections in the Hubei province with two schemes: fitting with the full datasets or fitting with the gradually shrinking datasets

| Treatments | $b$ | $\ln(a)$ | $M_0$ | $R$ | $p$-value | $N_{area}$ | $T_{day}$ |
|---|---|---|---|---|---|---|---|
| **Temporal Stability Model** | | | | | | | |
| Full Data between Jan 19 and Feb 29 | 2.292 | -2.990 | 10.109 | 0.996 | 0.000 | 17 | 40 |
| Minus tail 1 | 2.343 | -3.451 | 13.066 | 0.995 | 0.000 | 17 | 39 |
| Minus tail 2 | 2.392 | -3.920 | 16.695 | 0.993 | 0.000 | 17 | 38 |
| Minus tail 3 | 2.431 | -4.326 | 20.553 | 0.992 | 0.000 | 17 | 37 |
| Minus tail 4 | 2.445 | -4.575 | 23.704 | 0.991 | 0.000 | 17 | 36 |
| Minus tail 5 | 2.467 | -4.891 | 28.057 | 0.990 | 0.000 | 17 | 35 |
| Minus tail 6 | 2.505 | -5.332 | 34.587 | 0.988 | 0.000 | 17 | 34 |
| Minus tail 7 | 2.601 | -6.200 | 48.078 | 0.984 | 0.000 | 17 | 33 |
| Minus tail 8 | 2.584 | -6.285 | 52.909 | 0.983 | 0.000 | 17 | 32 |
| Minus tail 9 | 2.564 | -6.362 | 58.402 | 0.981 | 0.000 | 17 | 31 |
| Minus tail 10 | 2.541 | -6.431 | 64.926 | 0.979 | 0.000 | 17 | 30 |
| Minus tail 11 | 2.517 | -6.495 | 72.411 | 0.976 | 0.000 | 17 | 29 |
| Minus tail 12 | 2.489 | -6.559 | 81.739 | 0.972 | 0.000 | 17 | 28 |
| Minus tail 13 | 2.465 | -6.652 | 93.859 | 0.966 | 0.000 | 17 | 27 |
| Minus tail 14 | 2.443 | -6.777 | 109.376 | 0.961 | 0.000 | 17 | 26 |
| Minus tail 15 | 2.430 | -6.958 | 129.649 | 0.956 | 0.000 | 17 | 25 |
| | | | | | | | |
| Minus head 1 | 2.294 | -2.963 | 9.874 | 0.996 | 0.000 | 17 | 39 |
| Minus head 2 | 2.296 | -2.936 | 9.641 | 0.996 | 0.000 | 17 | 38 |
| Minus head 3 | 2.298 | -2.909 | 9.409 | 0.996 | 0.000 | 17 | 37 |
| Minus head 4 | 2.300 | -2.881 | 9.179 | 0.995 | 0.000 | 17 | 36 |
| Minus head 5 | 2.302 | -2.853 | 8.951 | 0.995 | 0.000 | 17 | 35 |
| Minus head 6 | 2.304 | -2.825 | 8.726 | 0.995 | 0.000 | 17 | 34 |
| Minus head 7 | 2.307 | -2.798 | 8.507 | 0.995 | 0.000 | 17 | 33 |
| Minus head 8 | 2.310 | -2.770 | 8.292 | 0.995 | 0.000 | 17 | 32 |
| Minus head 9 | 2.313 | -2.744 | 8.084 | 0.995 | 0.000 | 17 | 31 |
| Minus head 10 | 2.315 | -2.705 | 7.826 | 0.995 | 0.000 | 17 | 30 |
| Minus head 11 | 2.317 | -2.666 | 7.571 | 0.995 | 0.000 | 17 | 29 |
| Minus head 12 | 2.320 | -2.633 | 7.352 | 0.995 | 0.000 | 17 | 28 |
| Minus head 13 | 2.324 | -2.613 | 7.197 | 0.995 | 0.000 | 17 | 27 |
| Minus head 14 | 2.332 | -2.616 | 7.133 | 0.995 | 0.000 | 17 | 26 |
| Minus head 15 | 2.342 | -2.637 | 7.137 | 0.996 | 0.000 | 17 | 25 |

**Table S2A.** Fitting the TPL to the daily *incremental* COVID-19 infections in the whole China with two schemes: fitting with the full datasets or fitting with the gradually shrinking datasets

| Treatments | $b$ | $\ln(a)$ | $M_0$ | $R$ | $p$-value | $N_{area}$ | $T_{day}$ |
|---|---|---|---|---|---|---|---|
| **Temporal Stability Model** | | | | | | | |
| Full Data between Jan 19 and Feb 29 | 1.902 | 0.373 | 0.661 | 0.985 | 0.000 | 33 | 41 |
| Minus tail 1 | 1.881 | 0.412 | 0.626 | 0.983 | 0.000 | 33 | 40 |
| Minus tail 2 | 1.872 | 0.436 | 0.607 | 0.983 | 0.000 | 33 | 39 |
| Minus tail 3 | 1.872 | 0.450 | 0.597 | 0.984 | 0.000 | 33 | 38 |
| Minus tail 4 | 1.854 | 0.508 | 0.552 | 0.984 | 0.000 | 33 | 37 |
| Minus tail 5 | 1.817 | 0.619 | 0.469 | 0.982 | 0.000 | 33 | 36 |
| Minus tail 6 | 1.815 | 0.657 | 0.447 | 0.982 | 0.000 | 33 | 35 |
| Minus tail 7 | 1.820 | 0.662 | 0.446 | 0.983 | 0.000 | 33 | 34 |
| Minus tail 8 | 1.815 | 0.705 | 0.421 | 0.983 | 0.000 | 33 | 33 |
| Minus tail 9 | 1.799 | 0.767 | 0.383 | 0.983 | 0.000 | 33 | 32 |
| Minus tail 10 | 1.773 | 0.857 | 0.330 | 0.983 | 0.000 | 33 | 31 |
| Minus tail 11 | 1.842 | 0.739 | 0.416 | 0.982 | 0.000 | 32 | 30 |
| Minus tail 12 | 1.848 | 0.753 | 0.412 | 0.982 | 0.000 | 32 | 29 |
| Minus tail 13 | 1.984 | 0.470 | 0.620 | 0.991 | 0.000 | 31 | 28 |
| Minus tail 14 | 1.964 | 0.569 | 0.554 | 0.991 | 0.000 | 31 | 27 |
| Minus tail 15 | 1.938 | 0.665 | 0.492 | 0.988 | 0.000 | 31 | 26 |
| | | | | | | | |
| Minus head 1 | 1.900 | 0.333 | 0.691 | 0.984 | 0.000 | 33 | 40 |
| Minus head 2 | 1.901 | 0.281 | 0.732 | 0.983 | 0.000 | 33 | 39 |
| Minus head 3 | 1.898 | 0.242 | 0.763 | 0.983 | 0.000 | 33 | 38 |
| Minus head 4 | 1.898 | 0.197 | 0.803 | 0.982 | 0.000 | 33 | 37 |
| Minus head 5 | 1.894 | 0.151 | 0.845 | 0.981 | 0.000 | 33 | 36 |
| Minus head 6 | 1.890 | 0.104 | 0.889 | 0.980 | 0.000 | 33 | 35 |
| Minus head 7 | 1.882 | 0.062 | 0.932 | 0.979 | 0.000 | 33 | 34 |
| Minus head 8 | 1.876 | 0.011 | 0.988 | 0.977 | 0.000 | 33 | 33 |
| Minus head 9 | 1.876 | -0.057 | 1.067 | 0.976 | 0.000 | 33 | 32 |
| Minus head 10 | 1.851 | -0.121 | 1.153 | 0.975 | 0.000 | 33 | 31 |
| Minus head 11 | 1.843 | -0.176 | 1.232 | 0.973 | 0.000 | 33 | 30 |
| Minus head 12 | 1.835 | -0.229 | 1.315 | 0.970 | 0.000 | 33 | 29 |
| Minus head 13 | 1.826 | -0.279 | 1.402 | 0.967 | 0.000 | 33 | 28 |
| Minus head 14 | 1.825 | -0.350 | 1.528 | 0.965 | 0.000 | 33 | 27 |
| Minus head 15 | 1.822 | -0.405 | 1.637 | 0.962 | 0.000 | 33 | 26 |

**Table S2B.** Fitting the TPL to the daily *incremental* COVID-19 infections in the whole China except for Hubei province with two schemes: fitting with the full datasets or fitting with the gradually shrinking datasets

| Treatments | $b$ | $\ln(a)$ | $M_0$ | $R$ | $p$-value | $N_{area}$ | $T_{day}$ |
|---|---|---|---|---|---|---|---|
| **Temporal Stability Model** | | | | | | | |
| Full Data between Jan 19 and Feb 29 | 1.753 | 0.596 | 0.453 | 0.980 | 0.000 | 32 | 40 |
| Minus tail 1 | 1.727 | 0.641 | 0.414 | 0.978 | 0.000 | 32 | 39 |
| Minus tail 2 | 1.719 | 0.664 | 0.397 | 0.978 | 0.000 | 32 | 38 |
| Minus tail 3 | 1.722 | 0.673 | 0.394 | 0.979 | 0.000 | 32 | 37 |
| Minus tail 4 | 1.709 | 0.721 | 0.361 | 0.980 | 0.000 | 32 | 36 |
| Minus tail 5 | 1.669 | 0.827 | 0.290 | 0.978 | 0.000 | 32 | 35 |
| Minus tail 6 | 1.670 | 0.859 | 0.278 | 0.978 | 0.000 | 32 | 34 |
| Minus tail 7 | 1.680 | 0.855 | 0.284 | 0.978 | 0.000 | 32 | 33 |
| Minus tail 8 | 1.681 | 0.883 | 0.273 | 0.979 | 0.000 | 32 | 32 |
| Minus tail 9 | 1.665 | 0.935 | 0.245 | 0.978 | 0.000 | 32 | 31 |
| Minus tail 10 | 1.648 | 1.002 | 0.213 | 0.979 | 0.000 | 32 | 30 |
| Minus tail 11 | 1.717 | 0.906 | 0.282 | 0.972 | 0.000 | 31 | 29 |
| Minus tail 12 | 1.729 | 0.906 | 0.289 | 0.971 | 0.000 | 31 | 28 |
| Minus tail 13 | 1.947 | 0.522 | 0.576 | 0.981 | 0.000 | 30 | 27 |
| Minus tail 14 | 1.923 | 0.625 | 0.508 | 0.979 | 0.000 | 30 | 26 |
| Minus tail 15 | 1.870 | 0.751 | 0.422 | 0.973 | 0.000 | 30 | 25 |
| | | | | | | | |
| Minus head 1 | 1.747 | 0.566 | 0.468 | 0.979 | 0.000 | 32 | 39 |
| Minus head 2 | 1.746 | 0.521 | 0.498 | 0.979 | 0.000 | 32 | 38 |
| Minus head 3 | 1.738 | 0.493 | 0.513 | 0.978 | 0.000 | 32 | 37 |
| Minus head 4 | 1.734 | 0.457 | 0.536 | 0.978 | 0.000 | 32 | 36 |
| Minus head 5 | 1.724 | 0.424 | 0.557 | 0.977 | 0.000 | 32 | 35 |
| Minus head 6 | 1.714 | 0.392 | 0.577 | 0.976 | 0.000 | 32 | 34 |
| Minus head 7 | 1.700 | 0.365 | 0.594 | 0.975 | 0.000 | 32 | 33 |
| Minus head 8 | 1.686 | 0.332 | 0.616 | 0.972 | 0.000 | 32 | 32 |
| Minus head 9 | 1.678 | 0.282 | 0.660 | 0.971 | 0.000 | 32 | 31 |
| Minus head 10 | 1.625 | 0.272 | 0.647 | 0.974 | 0.000 | 32 | 30 |
| Minus head 11 | 1.608 | 0.239 | 0.674 | 0.972 | 0.000 | 32 | 29 |
| Minus head 12 | 1.587 | 0.216 | 0.692 | 0.969 | 0.000 | 32 | 28 |
| Minus head 13 | 1.564 | 0.195 | 0.708 | 0.967 | 0.000 | 32 | 27 |
| Minus head 14 | 1.553 | 0.153 | 0.759 | 0.966 | 0.000 | 32 | 26 |
| Minus head 15 | 1.540 | 0.121 | 0.799 | 0.963 | 0.000 | 32 | 25 |

**Table S2C.** Fitting the TPL to the daily *incremental* COVID-19 infections in Hubei province with two schemes: fitting with the full datasets or fitting with the gradually shrinking datasets

| Treatments | $b$ | $\ln(a)$ | $M_0$ | $R$ | $p$-value | $N_{area}$ | $T_{day}$ |
|---|---|---|---|---|---|---|---|
| **Temporal Stability Model** | | | | | | | |
| Full Data between Jan 19 and Feb 29 | 1.912 | 0.783 | 0.424 | 0.979 | 0.000 | 17 | 39 |
| Minus tail 1 | 1.877 | 0.888 | 0.363 | 0.978 | 0.000 | 17 | 38 |
| Minus tail 2 | 1.837 | 1.031 | 0.292 | 0.976 | 0.000 | 17 | 37 |
| Minus tail 3 | 1.784 | 1.231 | 0.208 | 0.974 | 0.000 | 17 | 36 |
| Minus tail 4 | 1.786 | 1.243 | 0.206 | 0.975 | 0.000 | 17 | 35 |
| Minus tail 5 | 1.786 | 1.264 | 0.200 | 0.976 | 0.000 | 17 | 34 |
| Minus tail 6 | 1.789 | 1.296 | 0.194 | 0.977 | 0.000 | 17 | 33 |
| Minus tail 7 | 1.753 | 1.461 | 0.144 | 0.981 | 0.000 | 17 | 32 |
| Minus tail 8 | 1.758 | 1.482 | 0.142 | 0.981 | 0.000 | 17 | 31 |
| Minus tail 9 | 1.763 | 1.513 | 0.138 | 0.982 | 0.000 | 17 | 30 |
| Minus tail 10 | 1.767 | 1.518 | 0.138 | 0.981 | 0.000 | 17 | 29 |
| Minus tail 11 | 1.773 | 1.553 | 0.134 | 0.982 | 0.000 | 17 | 28 |
| Minus tail 12 | 1.776 | 1.581 | 0.130 | 0.981 | 0.000 | 17 | 27 |
| Minus tail 13 | 1.776 | 1.599 | 0.128 | 0.979 | 0.000 | 17 | 26 |
| Minus tail 14 | 1.786 | 1.589 | 0.132 | 0.976 | 0.000 | 17 | 25 |
| Minus tail 15 | 1.797 | 1.605 | 0.134 | 0.974 | 0.000 | 17 | 24 |
| | | | | | | | |
| Minus head 1 | 1.914 | 0.732 | 0.449 | 0.978 | 0.000 | 17 | 38 |
| Minus head 2 | 1.916 | 0.681 | 0.476 | 0.977 | 0.000 | 17 | 37 |
| Minus head 3 | 1.917 | 0.629 | 0.504 | 0.976 | 0.000 | 17 | 36 |
| Minus head 4 | 1.919 | 0.572 | 0.537 | 0.975 | 0.000 | 17 | 35 |
| Minus head 5 | 1.922 | 0.512 | 0.574 | 0.973 | 0.000 | 17 | 34 |
| Minus head 6 | 1.924 | 0.450 | 0.614 | 0.972 | 0.000 | 17 | 33 |
| Minus head 7 | 1.926 | 0.390 | 0.656 | 0.970 | 0.000 | 17 | 32 |
| Minus head 8 | 1.929 | 0.325 | 0.705 | 0.969 | 0.000 | 17 | 31 |
| Minus head 9 | 1.906 | 0.413 | 0.634 | 0.965 | 0.000 | 17 | 30 |
| Minus head 10 | 1.907 | 0.356 | 0.675 | 0.963 | 0.000 | 17 | 29 |
| Minus head 11 | 1.919 | 0.149 | 0.850 | 0.965 | 0.000 | 17 | 28 |
| Minus head 12 | 1.930 | 0.016 | 0.983 | 0.963 | 0.000 | 17 | 27 |
| Minus head 13 | 1.939 | -0.062 | 1.069 | 0.961 | 0.000 | 17 | 26 |
| Minus head 14 | 1.940 | -0.120 | 1.136 | 0.957 | 0.000 | 17 | 25 |
| Minus head 15 | 1.991 | -0.429 | 1.542 | 0.958 | 0.000 | 17 | 24 |

**Table S3.** The randomization (permutation) test results of the TPL stability model parameters

| Treatments | Treatment A | Treatment B | TPL Parameter | p-value Cumulative | p-value Incremental |
|---|---|---|---|---|---|
| Whole Country | Raw data sets | Minus head-15 | $b$ | 0.781 | 0.760 |
| | | | $\ln(a)$ | 0.430 | 0.162 |
| | | | $M_0$ | 0.192 | 0.124 |
| | Raw data sets | Minus tail-15 | $b$ | 0.991 | 0.841 |
| | | | $\ln(a)$ | 0.131 | 0.345 |
| | | | $M_0$ | 0.000 | 0.500 |
| | Minus head-15 | Minus tail-15 | $b$ | 0.936 | 0.604 |
| | | | $\ln(a)$ | 0.084 | 0.018 |
| | | | $M_0$ | 0.001 | 0.009 |
| Whole Country Except Hubei | Raw data sets | Minus head-15 | $b$ | 0.816 | 0.161 |
| | | | $\ln(a)$ | 0.511 | 0.158 |
| | | | $M_0$ | 0.293 | 0.305 |
| | Raw data sets | Minus tail-15 | $b$ | 0.251 | 0.426 |
| | | | $\ln(a)$ | 0.438 | 0.573 |
| | | | $M_0$ | 0.000 | 0.880 |
| | Minus head-15 | Minus tail-15 | $b$ | 0.189 | 0.016 |
| | | | $\ln(a)$ | 0.255 | 0.018 |
| | | | $M_0$ | 0.002 | 0.077 |
| Hubei | Raw data sets | Minus head-15 | $b$ | 0.740 | 0.820 |
| | | | $\ln(a)$ | 0.718 | 0.369 |
| | | | $M_0$ | 0.489 | 0.344 |
| | Raw data sets | Minus tail-15 | $b$ | 0.784 | 0.715 |
| | | | $\ln(a)$ | 0.231 | 0.267 |
| | | | $M_0$ | 0.014 | 0.353 |
| | Minus head-15 | Minus tail-15 | $b$ | 0.898 | 0.604 |
| | | | $\ln(a)$ | 0.315 | 0.022 |
| | | | $M_0$ | 0.053 | 0.016 |
| Whole world | Raw data sets | Minus head-15 | $b$ | 0.490 | 0.956 |
| | | | $\ln(a)$ | 0.102 | 0.086 |
| | | | $M_0$ | 0.039 | 0.211 |
| | Raw data sets | Minus tail-15 | $b$ | 0.842 | 0.860 |
| | | | $\ln(a)$ | 0.202 | 0.381 |
| | | | $M_0$ | 0.065 | 0.513 |
| | Minus head-15 | Minus tail-15 | $b$ | 0.382 | 0.882 |
| | | | $\ln(a)$ | 0.715 | 0.011 |
| | | | $M_0$ | 0.826 | 0.038 |
| Whole World Except China | Raw data sets | Minus head-15 | $b$ | 0.259 | 0.245 |
| | | | $\ln(a)$ | 0.063 | 0.015 |
| | | | $M_0$ | 0.054 | 0.559 |
| | Raw data sets | Minus tail-15 | $b$ | 0.722 | 0.955 |
| | | | $\ln(a)$ | 0.204 | 0.362 |
| | | | $M_0$ | 0.049 | 0.618 |
| | Minus head-15 | Minus tail-15 | $b$ | 0.122 | 0.141 |
| | | | ln($a$) | 0.430 | 0.002 |
| | | | $M_0$ | 0.761 | 0.256 |
| Raw data sets | Whole Country | Whole Country (Except Hubei) | $b$ | 0.624 | 0.516 |
| | | | ln($a$) | 0.667 | 0.574 |
| | | | $M_0$ | 0.707 | 0.558 |
| | Whole Country | Hubei | $b$ | 0.197 | 0.966 |
| | | | ln($a$) | 0.139 | 0.472 |
| | | | $M_0$ | 0.064 | 0.606 |
| | Whole Country | Whole World | $b$ | 0.573 | 0.602 |
| | | | ln($a$) | 0.091 | 0.000 |
| | | | $M_0$ | 0.009 | 0.000 |
| | Whole Country | Whole World (Except China) | $b$ | 0.649 | 0.561 |
| | | | ln($a$) | 0.113 | 0.000 |
| | | | $M_0$ | 0.015 | 0.000 |
| | Whole Country (Except Hubei) | Hubei | $b$ | 0.157 | 0.529 |
| | | | ln($a$) | 0.117 | 0.766 |
| | | | $M_0$ | 0.047 | 0.966 |
| | Whole Country (Except Hubei) | Whole World | $b$ | 0.856 | 0.697 |
| | | | ln($a$) | 0.231 | 0.000 |
| | | | $M_0$ | 0.065 | 0.007 |
| | Whole Country (Except Hubei) | Whole World (Except China) | $b$ | 0.845 | 0.770 |
| | | | ln($a$) | 0.267 | 0.000 |
| | | | $M_0$ | 0.093 | 0.007 |
| | Hubei | Whole World | $b$ | 0.156 | 0.460 |
| | | | ln($a$) | 0.001 | 0.001 |
| | | | $M_0$ | 0.000 | 0.004 |
| | Hubei | Whole World (Except China) | $b$ | 0.200 | 0.412 |
| | | | ln($a$) | 0.002 | 0.003 |
| | | | $M_0$ | 0.000 | 0.005 |
| | Whole World | Whole World (Except Chinai | $b$ | 0.989 | 0.848 |
| | | | ln($a$) | 0.980 | 0.993 |
| | | | $M_0$ | 0.991 | 0.887 |
| Percentages (%) with Significant Differences in TPL Stability Parameters | | | $b$ | 0/25=0 | 1/25=4% |
| | | | ln($a$) | 2/25=8% | 12/25=48% |
| | | | $M_0$ | 12/25=48% | 9/25=36% |

**Table S4A.** The results of fitting TPL to the *cumulative* infections of COVID-19 worldwide with two schemes: fitting with the full datasets or fitting with the gradually shrinking datasets

| Treatments | $b$ | $\ln(a)$ | $M_0$ | $R$ | $p$-value | $N_{area}$ | $T_{day}$ |
|---|---|---|---|---|---|---|---|
| **Temporal Stability Model** | | | | | | | |
| Full Data between Jan 19 and Feb 29 | 2.076 | -1.069 | 2.701 | 0.944 | 0.000 | 37 | 41 |
| Minus tail 1 | 2.072 | -1.221 | 3.121 | 0.939 | 0.000 | 37 | 40 |
| Minus tail 2 | 2.013 | -1.355 | 3.810 | 0.906 | 0.000 | 35 | 39 |
| Minus tail 3 | 1.975 | -1.421 | 4.296 | 0.877 | 0.000 | 33 | 38 |
| Minus tail 4 | 2.004 | -1.617 | 5.000 | 0.879 | 0.000 | 28 | 37 |
| Minus tail 5 | 2.019 | -1.773 | 5.699 | 0.875 | 0.000 | 26 | 36 |
| Minus tail 6 | 2.246 | -1.800 | 4.237 | 0.966 | 0.000 | 22 | 35 |
| Minus tail 7 | 2.197 | -1.619 | 3.868 | 0.966 | 0.000 | 21 | 34 |
| Minus tail 8 | 2.185 | -1.641 | 3.993 | 0.965 | 0.000 | 21 | 33 |
| Minus tail 9 | 2.167 | -1.590 | 3.905 | 0.963 | 0.000 | 20 | 32 |
| Minus tail 10 | 2.165 | -1.663 | 4.168 | 0.961 | 0.000 | 20 | 31 |
| Minus tail 11 | 2.154 | -1.711 | 4.405 | 0.955 | 0.000 | 19 | 30 |
| Minus tail 12 | 2.148 | -1.765 | 4.654 | 0.951 | 0.000 | 19 | 29 |
| Minus tail 13 | 2.141 | -1.818 | 4.921 | 0.948 | 0.000 | 19 | 28 |
| Minus tail 14 | 2.136 | -1.894 | 5.296 | 0.942 | 0.000 | 19 | 27 |
| Minus tail 15 | 2.130 | -1.969 | 5.711 | 0.936 | 0.000 | 19 | 26 |
| | | | | | | | |
| Minus head 1 | 2.134 | -1.623 | 4.183 | 0.945 | 0.000 | 35 | 40 |
| Minus head 2 | 2.155 | -1.719 | 4.429 | 0.947 | 0.000 | 33 | 39 |
| Minus head 3 | 2.196 | -1.872 | 4.784 | 0.946 | 0.000 | 32 | 38 |
| Minus head 4 | 2.067 | -1.267 | 3.281 | 0.936 | 0.000 | 23 | 37 |
| Minus head 5 | 2.078 | -1.384 | 3.611 | 0.938 | 0.000 | 23 | 36 |
| Minus head 6 | 2.134 | -1.722 | 4.564 | 0.944 | 0.000 | 21 | 35 |
| Minus head 7 | 2.095 | -1.580 | 4.233 | 0.936 | 0.000 | 20 | 34 |
| Minus head 8 | 2.311 | -2.006 | 4.616 | 0.981 | 0.000 | 19 | 33 |
| Minus head 9 | 2.338 | -2.216 | 5.244 | 0.986 | 0.000 | 19 | 32 |
| Minus head 10 | 2.346 | -2.337 | 5.674 | 0.989 | 0.000 | 19 | 31 |
| Minus head 11 | 2.357 | -2.407 | 5.890 | 0.991 | 0.000 | 18 | 30 |
| Minus head 12 | 2.356 | -2.367 | 5.735 | 0.991 | 0.000 | 18 | 29 |
| Minus head 13 | 2.355 | -2.324 | 5.557 | 0.990 | 0.000 | 18 | 28 |
| Minus head 14 | 2.352 | -2.279 | 5.397 | 0.990 | 0.000 | 18 | 27 |
| Minus head 15 | 2.348 | -2.231 | 5.232 | 0.990 | 0.000 | 18 | 26 |

**Table S4B.** The results of fitting TPL to the *cumulative* infections of COVID-19 worldwide except for China with two schemes: fitting with the full datasets or fitting with the gradually shrinking datasets

| Treatments | $b$ | $\ln(a)$ | $M_0$ | $R$ | $p$-value | $N_{area}$ | $T_{day}$ |
|---|---|---|---|---|---|---|---|
| **Temporal Stability Model** | | | | | | | |
| Full Data between Jan 19 and Feb 29 | 2.069 | -1.055 | 2.683 | 0.910 | 0.000 | 36 | 41 |
| Minus tail 1 | 2.053 | -1.181 | 3.068 | 0.901 | 0.000 | 36 | 40 |
| Minus tail 2 | 1.929 | -1.170 | 3.525 | 0.845 | 0.000 | 34 | 39 |
| Minus tail 3 | 1.855 | -1.153 | 3.850 | 0.801 | 0.000 | 32 | 38 |
| Minus tail 4 | 1.881 | -1.306 | 4.403 | 0.794 | 0.000 | 27 | 37 |
| Minus tail 5 | 1.899 | -1.463 | 5.086 | 0.789 | 0.000 | 25 | 36 |
| Minus tail 6 | 2.439 | -2.285 | 4.896 | 0.948 | 0.000 | 21 | 35 |
| Minus tail 7 | 2.355 | -2.037 | 4.497 | 0.943 | 0.000 | 20 | 34 |
| Minus tail 8 | 2.331 | -2.029 | 4.594 | 0.942 | 0.000 | 20 | 33 |
| Minus tail 9 | 2.302 | -1.958 | 4.500 | 0.939 | 0.000 | 19 | 32 |
| Minus tail 10 | 2.294 | -2.019 | 4.758 | 0.936 | 0.000 | 19 | 31 |
| Minus tail 11 | 2.309 | -2.123 | 5.062 | 0.916 | 0.000 | 18 | 30 |
| Minus tail 12 | 2.297 | -2.165 | 5.307 | 0.910 | 0.000 | 18 | 29 |
| Minus tail 13 | 2.286 | -2.208 | 5.571 | 0.903 | 0.000 | 18 | 28 |
| Minus tail 14 | 2.276 | -2.274 | 5.944 | 0.893 | 0.000 | 18 | 27 |
| Minus tail 15 | 2.264 | -2.337 | 6.352 | 0.882 | 0.000 | 18 | 26 |
| | | | | | | | |
| Minus head 1 | 2.128 | -1.612 | 4.171 | 0.909 | 0.000 | 34 | 40 |
| Minus head 2 | 2.157 | -1.723 | 4.433 | 0.909 | 0.000 | 32 | 39 |
| Minus head 3 | 2.223 | -1.924 | 4.820 | 0.910 | 0.000 | 31 | 38 |
| Minus head 4 | 1.974 | -1.037 | 2.900 | 0.860 | 0.000 | 22 | 37 |
| Minus head 5 | 1.980 | -1.155 | 3.251 | 0.864 | 0.000 | 22 | 36 |
| Minus head 6 | 2.069 | -1.572 | 4.349 | 0.877 | 0.000 | 20 | 35 |
| Minus head 7 | 1.955 | -1.246 | 3.687 | 0.845 | 0.000 | 19 | 34 |
| Minus head 8 | 2.726 | -2.905 | 5.381 | 0.964 | 0.000 | 18 | 33 |
| Minus head 9 | 2.765 | -3.111 | 5.828 | 0.976 | 0.000 | 18 | 32 |
| Minus head 10 | 2.729 | -3.106 | 6.030 | 0.981 | 0.000 | 18 | 31 |
| Minus head 11 | 2.768 | -3.242 | 6.258 | 0.987 | 0.000 | 17 | 30 |
| Minus head 12 | 2.769 | -3.197 | 6.098 | 0.986 | 0.000 | 17 | 29 |
| Minus head 13 | 2.775 | -3.157 | 5.922 | 0.984 | 0.000 | 17 | 28 |
| Minus head 14 | 2.772 | -3.101 | 5.752 | 0.982 | 0.000 | 17 | 27 |
| Minus head 15 | 2.767 | -3.037 | 5.575 | 0.980 | 0.000 | 17 | 26 |

**Table S5A.** The results of fitting TPL to the *daily incremental* infections of COVID-19 worldwide with two schemes: fitting with the full datasets or fitting with the gradually shrinking datasets

| Treatments | b | ln(a) | $M_0$ | R | p-value | $N_{area}$ | $T_{day}$ |
|---|---|---|---|---|---|---|---|
| Temporal Stability Model | | | | | | | |
| Full Data between Jan 19 and Feb 29 | 1.820 | 1.598 | 0.142 | 0.980 | 0.000 | 36 | 40 |
| Minus tail 1 | 1.807 | 1.563 | 0.144 | 0.975 | 0.000 | 35 | 39 |
| Minus tail 2 | 1.830 | 1.641 | 0.138 | 0.984 | 0.000 | 33 | 38 |
| Minus tail 3 | 1.807 | 1.762 | 0.113 | 0.985 | 0.000 | 29 | 37 |
| Minus tail 4 | 1.828 | 1.751 | 0.121 | 0.986 | 0.000 | 26 | 36 |
| Minus tail 5 | 1.830 | 1.797 | 0.115 | 0.987 | 0.000 | 24 | 35 |
| Minus tail 6 | 1.848 | 1.762 | 0.125 | 0.989 | 0.000 | 22 | 34 |
| Minus tail 7 | 1.834 | 1.777 | 0.119 | 0.989 | 0.000 | 21 | 33 |
| Minus tail 8 | 1.832 | 1.810 | 0.114 | 0.991 | 0.000 | 20 | 32 |
| Minus tail 9 | 1.823 | 1.797 | 0.113 | 0.991 | 0.000 | 20 | 31 |
| Minus tail 10 | 1.820 | 1.809 | 0.110 | 0.990 | 0.000 | 19 | 30 |
| Minus tail 11 | 1.813 | 1.821 | 0.106 | 0.991 | 0.000 | 19 | 29 |
| Minus tail 12 | 1.807 | 1.827 | 0.104 | 0.992 | 0.000 | 19 | 28 |
| Minus tail 13 | 1.804 | 1.821 | 0.104 | 0.993 | 0.000 | 19 | 27 |
| Minus tail 14 | 1.807 | 1.789 | 0.109 | 0.993 | 0.000 | 19 | 26 |
| Minus tail 15 | 1.802 | 1.791 | 0.107 | 0.994 | 0.000 | 19 | 25 |
| | | | | | | | |
| Minus head 1 | 1.792 | 1.501 | 0.150 | 0.981 | 0.000 | 35 | 39 |
| Minus head 2 | 1.750 | 1.380 | 0.159 | 0.968 | 0.000 | 33 | 38 |
| Minus head 3 | 1.718 | 1.463 | 0.130 | 0.980 | 0.000 | 29 | 37 |
| Minus head 4 | 1.771 | 1.361 | 0.171 | 0.981 | 0.000 | 23 | 36 |
| Minus head 5 | 1.821 | 1.448 | 0.172 | 0.988 | 0.000 | 21 | 35 |
| Minus head 6 | 1.800 | 1.466 | 0.160 | 0.982 | 0.000 | 21 | 34 |
| Minus head 7 | 1.792 | 1.508 | 0.149 | 0.981 | 0.000 | 20 | 33 |
| Minus head 8 | 1.784 | 1.484 | 0.150 | 0.981 | 0.000 | 19 | 32 |
| Minus head 9 | 1.787 | 1.373 | 0.175 | 0.987 | 0.000 | 19 | 31 |
| Minus head 10 | 1.790 | 1.411 | 0.168 | 0.989 | 0.000 | 18 | 30 |
| Minus head 11 | 1.782 | 1.334 | 0.181 | 0.989 | 0.000 | 18 | 29 |
| Minus head 12 | 1.787 | 1.259 | 0.202 | 0.989 | 0.000 | 18 | 28 |
| Minus head 13 | 1.796 | 1.228 | 0.214 | 0.989 | 0.000 | 18 | 27 |
| Minus head 14 | 1.801 | 1.213 | 0.220 | 0.988 | 0.000 | 18 | 26 |
| Minus head 15 | 1.813 | 1.201 | 0.228 | 0.988 | 0.000 | 18 | 25 |

**Table S5B.** The results of fitting TPL to the *daily incremental* infections of COVID-19 worldwide except for China with two schemes: fitting with the full datasets or fitting with the gradually shrinking datasets

| Treatments | b | ln(a) | $M_0$ | R | p-value | $N_{area}$ | $T_{day}$ |
|---|---|---|---|---|---|---|---|
| **Temporal Stability Model** | | | | | | | |
| Full Data between Jan 19 and Feb 29 | 1.797 | 1.596 | 0.135 | 0.972 | 0.000 | 35 | 40 |
| Minus tail 1 | 1.779 | 1.561 | 0.135 | 0.965 | 0.000 | 34 | 39 |
| Minus tail 2 | 1.816 | 1.640 | 0.134 | 0.977 | 0.000 | 32 | 38 |
| Minus tail 3 | 1.793 | 1.758 | 0.109 | 0.978 | 0.000 | 28 | 37 |
| Minus tail 4 | 1.825 | 1.751 | 0.120 | 0.979 | 0.000 | 25 | 36 |
| Minus tail 5 | 1.831 | 1.797 | 0.115 | 0.981 | 0.000 | 23 | 35 |
| Minus tail 6 | 1.860 | 1.762 | 0.129 | 0.984 | 0.000 | 21 | 34 |
| Minus tail 7 | 1.842 | 1.777 | 0.121 | 0.984 | 0.000 | 20 | 33 |
| Minus tail 8 | 1.845 | 1.809 | 0.117 | 0.986 | 0.000 | 19 | 32 |
| Minus tail 9 | 1.831 | 1.797 | 0.115 | 0.987 | 0.000 | 19 | 31 |
| Minus tail 10 | 1.834 | 1.810 | 0.114 | 0.984 | 0.000 | 18 | 30 |
| Minus tail 11 | 1.824 | 1.821 | 0.110 | 0.985 | 0.000 | 18 | 29 |
| Minus tail 12 | 1.817 | 1.827 | 0.107 | 0.987 | 0.000 | 18 | 28 |
| Minus tail 13 | 1.811 | 1.821 | 0.106 | 0.988 | 0.000 | 18 | 27 |
| Minus tail 14 | 1.811 | 1.789 | 0.110 | 0.989 | 0.000 | 18 | 26 |
| Minus tail 15 | 1.803 | 1.791 | 0.108 | 0.990 | 0.000 | 18 | 25 |
| | | | | | | | |
| Minus head 1 | 1.745 | 1.481 | 0.137 | 0.972 | 0.000 | 34 | 39 |
| Minus head 2 | 1.671 | 1.349 | 0.134 | 0.951 | 0.000 | 32 | 38 |
| Minus head 3 | 1.636 | 1.416 | 0.108 | 0.972 | 0.000 | 28 | 37 |
| Minus head 4 | 1.680 | 1.354 | 0.137 | 0.970 | 0.000 | 22 | 36 |
| Minus head 5 | 1.771 | 1.430 | 0.157 | 0.978 | 0.000 | 20 | 35 |
| Minus head 6 | 1.737 | 1.436 | 0.143 | 0.968 | 0.000 | 20 | 34 |
| Minus head 7 | 1.727 | 1.476 | 0.131 | 0.965 | 0.000 | 19 | 33 |
| Minus head 8 | 1.676 | 1.414 | 0.124 | 0.957 | 0.000 | 18 | 32 |
| Minus head 9 | 1.658 | 1.270 | 0.145 | 0.969 | 0.000 | 18 | 31 |
| Minus head 10 | 1.650 | 1.269 | 0.142 | 0.971 | 0.000 | 17 | 30 |
| Minus head 11 | 1.599 | 1.153 | 0.146 | 0.973 | 0.000 | 17 | 29 |
| Minus head 12 | 1.575 | 1.053 | 0.160 | 0.972 | 0.000 | 17 | 28 |
| Minus head 13 | 1.584 | 1.032 | 0.171 | 0.971 | 0.000 | 17 | 27 |
| Minus head 14 | 1.575 | 1.010 | 0.172 | 0.969 | 0.000 | 17 | 26 |
| Minus head 15 | 1.585 | 1.005 | 0.179 | 0.966 | 0.000 | 17 | 25 |

**Table S6A**. Fitting the TPL to the *cumulative SARS* infection worldwide with two schemes: fitting with the full datasets or fitting with the gradually shrinking datasets

| Treatments | $b$ | $\ln(a)$ | $M_0$ | $R$ | $p$-value | $N$ |
|---|---|---|---|---|---|---|
| Full datasets | 2.020 | -2.293 | 9.475 | 0.965 | 0.000 | 23 |
| Minus tail 1 | 2.001 | -2.362 | 10.600 | 0.953 | 0.000 | 23 |
| Minus tail 2 | 2.007 | -2.470 | 11.618 | 0.952 | 0.000 | 23 |
| Minus tail 3 | 2.023 | -2.632 | 13.097 | 0.950 | 0.000 | 23 |
| Minus tail 4 | 2.023 | -2.655 | 13.412 | 0.960 | 0.000 | 22 |
| Minus tail 5 | 2.039 | -2.789 | 14.641 | 0.960 | 0.000 | 22 |
| Minus tail 6 | 2.035 | -2.845 | 15.602 | 0.958 | 0.000 | 21 |
| Minus tail 7 | 2.040 | -2.943 | 16.945 | 0.957 | 0.000 | 21 |
| Minus tail 8 | 2.073 | -3.186 | 19.458 | 0.959 | 0.000 | 21 |
| Minus tail 9 | 2.083 | -3.308 | 21.186 | 0.958 | 0.000 | 21 |
| Minus tail 10 | 2.109 | -3.500 | 23.507 | 0.955 | 0.000 | 21 |
| Minus tail 11 | 2.109 | -3.617 | 26.124 | 0.947 | 0.000 | 20 |
| Minus tail 12 | 2.143 | -3.873 | 29.640 | 0.942 | 0.000 | 20 |
| Minus tail 13 | 2.122 | -3.862 | 31.277 | 0.934 | 0.000 | 19 |
| Minus tail 14 | 2.090 | -3.784 | 32.186 | 0.928 | 0.000 | 18 |
| Minus tail 15 | 2.055 | -3.733 | 34.388 | 0.910 | 0.000 | 17 |
| Minus tail 16 | 2.052 | -3.843 | 38.563 | 0.898 | 0.000 | 17 |
| Minus tail 17 | 2.036 | -3.634 | 33.344 | 0.912 | 0.000 | 14 |
| Minus tail 18 | 2.006 | -3.683 | 38.847 | 0.884 | 0.000 | 14 |
| Minus tail 19 | 1.997 | -3.722 | 41.885 | 0.881 | 0.000 | 14 |
| Minus tail 20 | 1.988 | -3.788 | 46.303 | 0.875 | 0.000 | 14 |
| Minus tail 21 | 1.970 | -3.794 | 49.900 | 0.868 | 0.000 | 14 |
| Minus tail 22 | 1.950 | -3.799 | 54.598 | 0.858 | 0.000 | 14 |
| Minus tail 23 | 1.924 | -3.814 | 62.126 | 0.840 | 0.000 | 14 |
| Minus tail 24 | 2.058 | -3.632 | 30.969 | 0.969 | 0.000 | 13 |
| Minus tail 25 | 2.043 | -3.643 | 32.879 | 0.967 | 0.000 | 13 |
| Minus tail 26 | 2.027 | -3.663 | 35.416 | 0.966 | 0.000 | 13 |
| Minus tail 27 | 2.004 | -3.693 | 39.599 | 0.965 | 0.000 | 13 |
| Minus tail 28 | 1.966 | -3.729 | 47.505 | 0.966 | 0.000 | 13 |
| Minus tail 29 | 1.950 | -3.726 | 50.579 | 0.964 | 0.000 | 13 |
| Minus tail 30 | 1.930 | -3.719 | 54.606 | 0.961 | 0.000 | 13 |
| Minus tail 31 | 1.910 | -3.721 | 59.666 | 0.958 | 0.000 | 13 |
| Minus tail 32 | 1.890 | -3.710 | 64.503 | 0.956 | 0.000 | 13 |
| Minus tail 33 | 1.873 | -3.706 | 69.866 | 0.953 | 0.000 | 13 |
| Minus tail 34 | 1.855 | -3.702 | 75.854 | 0.951 | 0.000 | 13 |
| Minus tail 35 | 1.839 | -3.707 | 82.985 | 0.949 | 0.000 | 13 |
| Minus tail 36 | 1.824 | -3.728 | 92.188 | 0.947 | 0.000 | 13 |
| Minus tail 37 | 1.814 | -3.768 | 102.489 | 0.943 | 0.000 | 13 |
| Minus tail 38 | 1.815 | -3.872 | 115.909 | 0.936 | 0.000 | 13 |
| Minus tail 39 | 1.807 | -3.918 | 128.464 | 0.932 | 0.000 | 13 |
| Minus tail 40 | 1.808 | -4.033 | 147.457 | 0.925 | 0.000 | 13 |
| Minus head 1 | 2.021 | -2.293 | 9.446 | 0.965 | 0.000 | 23 |
| Minus head 2 | 2.024 | -2.295 | 9.410 | 0.965 | 0.000 | 23 |
| Minus head 3 | 2.024 | -2.287 | 9.326 | 0.965 | 0.000 | 23 |
| Minus head 4 | 2.025 | -2.279 | 9.242 | 0.965 | 0.000 | 23 |
| Minus head 5 | 2.025 | -2.271 | 9.161 | 0.965 | 0.000 | 23 |
| Minus head 6 | 2.026 | -2.264 | 9.081 | 0.965 | 0.000 | 23 |
| Minus head 7 | 2.027 | -2.257 | 9.004 | 0.965 | 0.000 | 23 |
| Minus head 8 | 2.028 | -2.250 | 8.930 | 0.965 | 0.000 | 23 |
| Minus head 9 | 2.029 | -2.244 | 8.860 | 0.965 | 0.000 | 23 |
| Minus head 10 | 2.030 | -2.238 | 8.791 | 0.965 | 0.000 | 23 |
| Minus head 11 | 2.030 | -2.230 | 8.710 | 0.965 | 0.000 | 23 |
| Minus head 12 | 2.029 | -2.214 | 8.601 | 0.965 | 0.000 | 23 |
| Minus head 13 | 2.026 | -2.188 | 8.443 | 0.964 | 0.000 | 23 |
| Minus head 14 | 2.021 | -2.158 | 8.276 | 0.964 | 0.000 | 23 |
| Minus head 15 | 2.019 | -2.136 | 8.127 | 0.963 | 0.000 | 23 |
| Minus head 16 | 2.027 | -2.157 | 8.173 | 0.964 | 0.000 | 23 |
| Minus head 17 | 2.039 | -2.221 | 8.468 | 0.965 | 0.000 | 22 |
| Minus head 18 | 2.039 | -2.210 | 8.381 | 0.965 | 0.000 | 22 |
| Minus head 19 | 2.039 | -2.199 | 8.298 | 0.965 | 0.000 | 22 |
| Minus head 20 | 2.039 | -2.190 | 8.219 | 0.965 | 0.000 | 22 |
| Minus head 21 | 2.040 | -2.180 | 8.144 | 0.965 | 0.000 | 22 |
| Minus head 22 | 2.040 | -2.172 | 8.074 | 0.965 | 0.000 | 22 |
| Minus head 23 | 2.041 | -2.165 | 8.011 | 0.964 | 0.000 | 22 |
| Minus head 24 | 2.042 | -2.160 | 7.953 | 0.964 | 0.000 | 22 |
| Minus head 25 | 2.043 | -2.158 | 7.908 | 0.963 | 0.000 | 22 |
| Minus head 26 | 2.042 | -2.140 | 7.800 | 0.963 | 0.000 | 22 |
| Minus head 27 | 2.040 | -2.121 | 7.687 | 0.963 | 0.000 | 22 |
| Minus head 28 | 2.038 | -2.102 | 7.574 | 0.963 | 0.000 | 22 |
| Minus head 29 | 2.036 | -2.082 | 7.461 | 0.963 | 0.000 | 22 |
| Minus head 30 | 2.034 | -2.062 | 7.345 | 0.963 | 0.000 | 22 |
| Minus head 31 | 2.032 | -2.043 | 7.231 | 0.963 | 0.000 | 22 |
| Minus head 32 | 2.031 | -2.022 | 7.109 | 0.963 | 0.000 | 22 |
| Minus head 33 | 2.030 | -2.001 | 6.986 | 0.962 | 0.000 | 22 |
| Minus head 34 | 2.028 | -1.980 | 6.864 | 0.962 | 0.000 | 22 |
| Minus head 35 | 2.026 | -1.959 | 6.743 | 0.962 | 0.000 | 22 |
| Minus head 36 | 2.024 | -1.937 | 6.623 | 0.962 | 0.000 | 22 |
| Minus head 37 | 2.023 | -1.915 | 6.508 | 0.962 | 0.000 | 22 |
| Minus head 38 | 2.021 | -1.894 | 6.395 | 0.962 | 0.000 | 22 |
| Minus head 39 | 2.019 | -1.873 | 6.286 | 0.962 | 0.000 | 22 |
| Minus head 40 | 2.017 | -1.853 | 6.182 | 0.962 | 0.000 | 22 |

**Table S6B**. Fitting the TPL to the *cumulative SARS* infection worldwide except for China with two schemes: fitting with the full datasets or fitting with the gradually shrinking datasets

| Treatments | b | ln(a) | $M_0$ | R | p-value | N |
|---|---|---|---|---|---|---|
| Full datasets | 1.902 | -2.112 | 10.409 | 0.931 | 0.000 | 22 |
| Minus tail 1 | 1.857 | -2.142 | 12.174 | 0.909 | 0.000 | 22 |
| Minus tail 2 | 1.864 | -2.251 | 13.522 | 0.908 | 0.000 | 22 |
| Minus tail 3 | 1.885 | -2.419 | 15.397 | 0.904 | 0.000 | 22 |
| Minus tail 4 | 1.888 | -2.445 | 15.682 | 0.922 | 0.000 | 21 |
| Minus tail 5 | 1.914 | -2.593 | 17.057 | 0.924 | 0.000 | 21 |
| Minus tail 6 | 1.898 | -2.615 | 18.394 | 0.917 | 0.000 | 20 |
| Minus tail 7 | 1.904 | -2.715 | 20.177 | 0.916 | 0.000 | 20 |
| Minus tail 8 | 1.955 | -2.989 | 22.861 | 0.921 | 0.000 | 20 |
| Minus tail 9 | 1.968 | -3.115 | 24.991 | 0.918 | 0.000 | 20 |
| Minus tail 10 | 2.005 | -3.327 | 27.392 | 0.914 | 0.000 | 20 |
| Minus tail 11 | 1.987 | -3.400 | 31.314 | 0.895 | 0.000 | 19 |
| Minus tail 12 | 2.037 | -3.685 | 34.866 | 0.887 | 0.000 | 19 |
| Minus tail 13 | 1.997 | -3.633 | 38.234 | 0.871 | 0.000 | 18 |
| Minus tail 14 | 1.937 | -3.490 | 41.436 | 0.856 | 0.000 | 17 |
| Minus tail 15 | 1.845 | -3.295 | 49.334 | 0.813 | 0.000 | 16 |
| Minus tail 16 | 1.830 | -3.381 | 58.797 | 0.790 | 0.000 | 16 |
| Minus tail 17 | 1.794 | -3.061 | 47.312 | 0.802 | 0.001 | 13 |
| Minus tail 18 | 1.707 | -2.976 | 67.114 | 0.742 | 0.004 | 13 |
| Minus tail 19 | 1.683 | -2.980 | 78.415 | 0.733 | 0.004 | 13 |
| Minus tail 20 | 1.656 | -3.001 | 97.064 | 0.720 | 0.005 | 13 |
| Minus tail 21 | 1.618 | -2.957 | 119.885 | 0.704 | 0.007 | 13 |
| Minus tail 22 | 1.571 | -2.899 | 160.220 | 0.681 | 0.010 | 13 |
| Minus tail 23 | 1.506 | -2.821 | 263.037 | 0.643 | 0.018 | 13 |
| Minus tail 24 | 1.959 | -3.408 | 34.896 | 0.923 | 0.000 | 12 |
| Minus tail 25 | 1.930 | -3.386 | 38.153 | 0.919 | 0.000 | 12 |
| Minus tail 26 | 1.896 | -3.365 | 42.806 | 0.915 | 0.000 | 12 |
| Minus tail 27 | 1.838 | -3.317 | 52.228 | 0.911 | 0.000 | 12 |
| Minus tail 28 | 1.733 | -3.200 | 78.533 | 0.912 | 0.000 | 12 |
| Minus tail 29 | 1.714 | -3.188 | 87.252 | 0.908 | 0.000 | 12 |
| Minus tail 30 | 1.687 | -3.166 | 100.257 | 0.901 | 0.000 | 12 |
| Minus tail 31 | 1.656 | -3.142 | 120.081 | 0.893 | 0.000 | 12 |
| Minus tail 32 | 1.633 | -3.123 | 138.830 | 0.886 | 0.000 | 12 |
| Minus tail 33 | 1.614 | -3.115 | 160.279 | 0.879 | 0.000 | 12 |
| Minus tail 34 | 1.598 | -3.116 | 182.892 | 0.874 | 0.000 | 12 |
| Minus tail 35 | 1.587 | -3.132 | 207.718 | 0.869 | 0.000 | 12 |
| Minus tail 36 | 1.581 | -3.173 | 235.971 | 0.863 | 0.000 | 12 |
| Minus tail 37 | 1.588 | -3.253 | 252.254 | 0.856 | 0.000 | 12 |
| Minus tail 38 | 1.611 | -3.407 | 264.496 | 0.841 | 0.001 | 12 |
| Minus tail 39 | 1.623 | -3.499 | 274.375 | 0.833 | 0.001 | 12 |
| Minus tail 40 | 1.649 | -3.671 | 286.690 | 0.819 | 0.001 | 12 |
| Minus head 1 | 1.903 | -2.112 | 10.370 | 0.932 | 0.000 | 22 |
| Minus head 2 | 1.906 | -2.115 | 10.314 | 0.932 | 0.000 | 22 |
| Minus head 3 | 1.907 | -2.107 | 10.210 | 0.932 | 0.000 | 22 |
| Minus head 4 | 1.907 | -2.099 | 10.108 | 0.932 | 0.000 | 22 |
| Minus head 5 | 1.908 | -2.091 | 10.007 | 0.932 | 0.000 | 22 |
| Minus head 6 | 1.908 | -2.083 | 9.909 | 0.932 | 0.000 | 22 |
| Minus head 7 | 1.909 | -2.076 | 9.813 | 0.932 | 0.000 | 22 |
| Minus head 8 | 1.910 | -2.069 | 9.721 | 0.932 | 0.000 | 22 |
| Minus head 9 | 1.911 | -2.063 | 9.633 | 0.932 | 0.000 | 22 |
| Minus head 10 | 1.912 | -2.057 | 9.547 | 0.932 | 0.000 | 22 |
| Minus head 11 | 1.912 | -2.048 | 9.448 | 0.932 | 0.000 | 22 |
| Minus head 12 | 1.909 | -2.031 | 9.333 | 0.931 | 0.000 | 22 |
| Minus head 13 | 1.904 | -2.003 | 9.163 | 0.930 | 0.000 | 22 |
| Minus head 14 | 1.897 | -1.970 | 8.988 | 0.929 | 0.000 | 22 |
| Minus head 15 | 1.894 | -1.946 | 8.815 | 0.928 | 0.000 | 22 |
| Minus head 16 | 1.904 | -1.970 | 8.837 | 0.930 | 0.000 | 22 |
| Minus head 17 | 1.920 | -2.037 | 9.142 | 0.932 | 0.000 | 21 |
| Minus head 18 | 1.920 | -2.026 | 9.044 | 0.932 | 0.000 | 21 |
| Minus head 19 | 1.919 | -2.015 | 8.950 | 0.932 | 0.000 | 21 |
| Minus head 20 | 1.919 | -2.004 | 8.861 | 0.932 | 0.000 | 21 |
| Minus head 21 | 1.918 | -1.994 | 8.777 | 0.931 | 0.000 | 21 |
| Minus head 22 | 1.918 | -1.986 | 8.699 | 0.931 | 0.000 | 21 |
| Minus head 23 | 1.918 | -1.979 | 8.629 | 0.930 | 0.000 | 21 |
| Minus head 24 | 1.919 | -1.974 | 8.561 | 0.929 | 0.000 | 21 |
| Minus head 25 | 1.921 | -1.973 | 8.508 | 0.928 | 0.000 | 21 |
| Minus head 26 | 1.918 | -1.952 | 8.391 | 0.928 | 0.000 | 21 |
| Minus head 27 | 1.915 | -1.932 | 8.266 | 0.928 | 0.000 | 21 |
| Minus head 28 | 1.911 | -1.910 | 8.142 | 0.928 | 0.000 | 21 |
| Minus head 29 | 1.907 | -1.889 | 8.018 | 0.928 | 0.000 | 21 |
| Minus head 30 | 1.904 | -1.868 | 7.889 | 0.928 | 0.000 | 21 |
| Minus head 31 | 1.901 | -1.847 | 7.761 | 0.927 | 0.000 | 21 |
| Minus head 32 | 1.899 | -1.826 | 7.620 | 0.927 | 0.000 | 21 |
| Minus head 33 | 1.897 | -1.806 | 7.479 | 0.927 | 0.000 | 21 |
| Minus head 34 | 1.895 | -1.785 | 7.338 | 0.927 | 0.000 | 21 |
| Minus head 35 | 1.893 | -1.763 | 7.199 | 0.927 | 0.000 | 21 |
| Minus head 36 | 1.891 | -1.742 | 7.062 | 0.927 | 0.000 | 21 |
| Minus head 37 | 1.889 | -1.721 | 6.931 | 0.927 | 0.000 | 21 |
| Minus head 38 | 1.887 | -1.700 | 6.804 | 0.927 | 0.000 | 21 |
| Minus head 39 | 1.884 | -1.680 | 6.683 | 0.927 | 0.000 | 21 |
| Minus head 40 | 1.882 | -1.661 | 6.568 | 0.927 | 0.000 | 21 |

**Table S7A.** Fitting the TPL to the *daily incremental SARS* infection worldwide with two schemes: fitting with the full datasets or fitting with the gradually shrinking datasets

| Treatments | b | ln(a) | $M_0$ | R | p-value | N |
|---|---|---|---|---|---|---|
| Full datasets | 1.546 | 2.829 | 0.006 | 0.906 | 0.000 | 18 |
| Minus tail 1 | 1.509 | 2.817 | 0.004 | 0.897 | 0.000 | 18 |
| Minus tail 2 | 1.498 | 2.796 | 0.004 | 0.893 | 0.000 | 18 |
| Minus tail 3 | 1.341 | 2.669 | 0.000 | 0.836 | 0.000 | 18 |
| Minus tail 4 | 1.351 | 2.347 | 0.001 | 0.935 | 0.000 | 16 |
| Minus tail 5 | 1.376 | 2.327 | 0.002 | 0.948 | 0.000 | 15 |
| Minus tail 6 | 1.381 | 2.326 | 0.002 | 0.948 | 0.000 | 16 |
| Minus tail 7 | 1.383 | 2.343 | 0.002 | 0.949 | 0.000 | 16 |
| Minus tail 8 | 1.329 | 2.207 | 0.001 | 0.935 | 0.000 | 16 |
| Minus tail 9 | 1.393 | 2.200 | 0.004 | 0.949 | 0.000 | 15 |
| Minus tail 10 | 1.396 | 2.205 | 0.004 | 0.949 | 0.000 | 15 |
| Minus tail 11 | 1.403 | 2.268 | 0.004 | 0.952 | 0.000 | 15 |
| Minus tail 12 | 1.402 | 2.318 | 0.003 | 0.954 | 0.000 | 15 |
| Minus tail 13 | 1.402 | 2.398 | 0.003 | 0.957 | 0.000 | 14 |
| Minus tail 14 | 1.375 | 2.427 | 0.002 | 0.956 | 0.000 | 13 |
| Minus tail 15 | 1.377 | 2.447 | 0.002 | 0.957 | 0.000 | 13 |
| Minus tail 16 | 1.420 | 2.438 | 0.003 | 0.963 | 0.000 | 12 |
| Minus tail 17 | 1.388 | 2.503 | 0.002 | 0.961 | 0.000 | 10 |
| Minus tail 18 | 1.415 | 2.520 | 0.002 | 0.964 | 0.000 | 10 |
| Minus tail 19 | 1.418 | 2.529 | 0.002 | 0.964 | 0.000 | 10 |
| Minus tail 20 | 1.430 | 2.502 | 0.003 | 0.963 | 0.000 | 10 |
| Minus tail 21 | 1.434 | 2.537 | 0.003 | 0.964 | 0.000 | 10 |
| Minus tail 22 | 1.438 | 2.550 | 0.003 | 0.964 | 0.000 | 10 |
| Minus tail 23 | 1.417 | 2.594 | 0.002 | 0.961 | 0.000 | 10 |
| Minus tail 24 | 1.391 | 2.647 | 0.001 | 0.959 | 0.000 | 9 |
| Minus tail 25 | 1.393 | 2.653 | 0.001 | 0.959 | 0.000 | 9 |
| Minus tail 26 | 1.396 | 2.655 | 0.001 | 0.959 | 0.000 | 9 |
| Minus tail 27 | 1.404 | 2.705 | 0.001 | 0.961 | 0.000 | 9 |
| Minus tail 28 | 1.607 | 2.482 | 0.017 | 0.998 | 0.000 | 7 |
| Minus tail 29 | 1.606 | 2.470 | 0.017 | 0.998 | 0.000 | 7 |
| Minus tail 30 | 1.596 | 2.437 | 0.017 | 0.997 | 0.000 | 7 |
| Minus tail 31 | 1.587 | 2.518 | 0.014 | 0.997 | 0.000 | 7 |
| Minus tail 32 | 1.588 | 2.522 | 0.014 | 0.996 | 0.000 | 7 |
| Minus tail 33 | 1.548 | 2.596 | 0.009 | 0.993 | 0.000 | 7 |
| Minus tail 34 | 1.553 | 2.613 | 0.009 | 0.992 | 0.000 | 7 |
| Minus tail 35 | 1.639 | 2.521 | 0.019 | 0.994 | 0.000 | 7 |
| Minus tail 36 | 1.645 | 2.529 | 0.020 | 0.994 | 0.000 | 7 |
| Minus tail 37 | 1.636 | 2.508 | 0.019 | 0.993 | 0.000 | 7 |
| Minus tail 38 | 1.625 | 2.557 | 0.017 | 0.994 | 0.000 | 7 |
| Minus tail 39 | 1.629 | 2.553 | 0.017 | 0.994 | 0.000 | 7 |
| Minus tail 40 | 1.637 | 2.575 | 0.017 | 0.993 | 0.000 | 7 |
| Minus head 1 | 1.575 | 2.765 | 0.008 | 0.940 | 0.000 | 18 |
| Minus head 2 | 1.575 | 2.756 | 0.008 | 0.940 | 0.000 | 18 |
| Minus head 3 | 1.575 | 2.749 | 0.008 | 0.940 | 0.000 | 18 |
| Minus head 4 | 1.574 | 2.739 | 0.008 | 0.939 | 0.000 | 18 |
| Minus head 5 | 1.572 | 2.729 | 0.008 | 0.939 | 0.000 | 18 |
| Minus head 6 | 1.572 | 2.720 | 0.009 | 0.938 | 0.000 | 18 |
| Minus head 7 | 1.571 | 2.711 | 0.009 | 0.938 | 0.000 | 18 |
| Minus head 8 | 1.570 | 2.700 | 0.009 | 0.938 | 0.000 | 18 |
| Minus head 9 | 1.584 | 2.689 | 0.010 | 0.941 | 0.000 | 18 |
| Minus head 10 | 1.597 | 2.671 | 0.011 | 0.943 | 0.000 | 19 |
| Minus head 11 | 1.595 | 2.726 | 0.010 | 0.940 | 0.000 | 20 |
| Minus head 12 | 1.618 | 2.602 | 0.015 | 0.948 | 0.000 | 19 |
| Minus head 13 | 1.619 | 2.591 | 0.015 | 0.949 | 0.000 | 19 |
| Minus head 14 | 1.619 | 2.581 | 0.015 | 0.949 | 0.000 | 19 |
| Minus head 15 | 1.619 | 2.573 | 0.016 | 0.948 | 0.000 | 19 |
| Minus head 16 | 1.619 | 2.563 | 0.016 | 0.948 | 0.000 | 19 |
| Minus head 17 | 1.618 | 2.550 | 0.016 | 0.948 | 0.000 | 19 |
| Minus head 18 | 1.617 | 2.537 | 0.016 | 0.948 | 0.000 | 19 |
| Minus head 19 | 1.616 | 2.525 | 0.017 | 0.947 | 0.000 | 19 |
| Minus head 20 | 1.616 | 2.518 | 0.017 | 0.946 | 0.000 | 19 |
| Minus head 21 | 1.616 | 2.507 | 0.017 | 0.945 | 0.000 | 19 |
| Minus head 22 | 1.616 | 2.500 | 0.017 | 0.945 | 0.000 | 19 |
| Minus head 23 | 1.619 | 2.505 | 0.018 | 0.945 | 0.000 | 19 |
| Minus head 24 | 1.618 | 2.495 | 0.018 | 0.944 | 0.000 | 19 |
| Minus head 25 | 1.608 | 2.490 | 0.017 | 0.945 | 0.000 | 19 |
| Minus head 26 | 1.609 | 2.491 | 0.017 | 0.945 | 0.000 | 19 |
| Minus head 27 | 1.609 | 2.477 | 0.017 | 0.945 | 0.000 | 19 |
| Minus head 28 | 1.609 | 2.469 | 0.017 | 0.944 | 0.000 | 19 |
| Minus head 29 | 1.610 | 2.462 | 0.018 | 0.943 | 0.000 | 19 |
| Minus head 30 | 1.617 | 2.436 | 0.019 | 0.942 | 0.000 | 19 |
| Minus head 31 | 1.613 | 2.413 | 0.019 | 0.941 | 0.000 | 19 |
| Minus head 32 | 1.615 | 2.406 | 0.020 | 0.941 | 0.000 | 19 |
| Minus head 33 | 1.617 | 2.396 | 0.021 | 0.941 | 0.000 | 19 |
| Minus head 34 | 1.616 | 2.383 | 0.021 | 0.940 | 0.000 | 19 |
| Minus head 35 | 1.616 | 2.367 | 0.021 | 0.939 | 0.000 | 19 |
| Minus head 36 | 1.633 | 2.301 | 0.026 | 0.940 | 0.000 | 19 |
| Minus head 37 | 1.633 | 2.284 | 0.027 | 0.940 | 0.000 | 19 |
| Minus head 38 | 1.634 | 2.269 | 0.028 | 0.939 | 0.000 | 19 |
| Minus head 39 | 1.636 | 2.254 | 0.029 | 0.940 | 0.000 | 19 |
| Minus head 40 | 1.637 | 2.245 | 0.029 | 0.940 | 0.000 | 19 |

**Table S7B.** Fitting the TPL to the *daily incremental SARS* infection worldwide except for China with two schemes: fitting with the full datasets or fitting with the gradually shrinking datasets

| Treatments | b | ln(a) | $M_0$ | R | p-value | N |
|---|---|---|---|---|---|---|
| Full datasets | 1.537 | 2.806 | 0.005 | 0.815 | 0.000 | 17 |
| Minus tail 1 | 1.466 | 2.698 | 0.003 | 0.792 | 0.000 | 17 |
| Minus tail 2 | 1.442 | 2.639 | 0.003 | 0.782 | 0.000 | 17 |
| Minus tail 3 | 1.142 | 2.080 | 0.000 | 0.667 | 0.003 | 17 |
| Minus tail 4 | 1.084 | 1.554 | 0.000 | 0.844 | 0.000 | 15 |
| Minus tail 5 | 1.130 | 1.599 | 0.000 | 0.881 | 0.000 | 14 |
| Minus tail 6 | 1.133 | 1.584 | 0.000 | 0.877 | 0.000 | 15 |
| Minus tail 7 | 1.139 | 1.609 | 0.000 | 0.879 | 0.000 | 15 |
| Minus tail 8 | 1.085 | 1.460 | 0.000 | 0.846 | 0.000 | 15 |
| Minus tail 9 | 1.184 | 1.582 | 0.000 | 0.878 | 0.000 | 14 |
| Minus tail 10 | 1.187 | 1.588 | 0.000 | 0.876 | 0.000 | 14 |
| Minus tail 11 | 1.213 | 1.695 | 0.000 | 0.885 | 0.000 | 14 |
| Minus tail 12 | 1.221 | 1.757 | 0.000 | 0.888 | 0.000 | 14 |
| Minus tail 13 | 1.234 | 1.867 | 0.000 | 0.897 | 0.000 | 13 |
| Minus tail 14 | 1.195 | 1.867 | 0.000 | 0.895 | 0.000 | 12 |
| Minus tail 15 | 1.201 | 1.901 | 0.000 | 0.896 | 0.000 | 12 |
| Minus tail 16 | 1.270 | 1.990 | 0.001 | 0.909 | 0.000 | 11 |
| Minus tail 17 | 1.208 | 2.000 | 0.000 | 0.904 | 0.001 | 9 |
| Minus tail 18 | 1.272 | 2.092 | 0.000 | 0.912 | 0.001 | 9 |
| Minus tail 19 | 1.277 | 2.108 | 0.000 | 0.911 | 0.001 | 9 |
| Minus tail 20 | 1.285 | 2.058 | 0.001 | 0.905 | 0.001 | 9 |
| Minus tail 21 | 1.298 | 2.118 | 0.001 | 0.907 | 0.001 | 9 |
| Minus tail 22 | 1.304 | 2.139 | 0.001 | 0.906 | 0.001 | 9 |
| Minus tail 23 | 1.271 | 2.136 | 0.000 | 0.900 | 0.001 | 9 |
| Minus tail 24 | 1.207 | 2.104 | 0.000 | 0.889 | 0.003 | 8 |
| Minus tail 25 | 1.211 | 2.117 | 0.000 | 0.888 | 0.003 | 8 |
| Minus tail 26 | 1.214 | 2.123 | 0.000 | 0.887 | 0.003 | 8 |
| Minus tail 27 | 1.233 | 2.190 | 0.000 | 0.888 | 0.003 | 8 |
| Minus tail 28 | 1.616 | 2.505 | 0.017 | 0.993 | 0.000 | 6 |
| Minus tail 29 | 1.616 | 2.494 | 0.017 | 0.993 | 0.000 | 6 |
| Minus tail 30 | 1.619 | 2.494 | 0.018 | 0.993 | 0.000 | 6 |
| Minus tail 31 | 1.618 | 2.600 | 0.015 | 0.990 | 0.000 | 6 |
| Minus tail 32 | 1.630 | 2.631 | 0.015 | 0.989 | 0.000 | 6 |
| Minus tail 33 | 1.543 | 2.582 | 0.009 | 0.979 | 0.001 | 6 |
| Minus tail 34 | 1.550 | 2.605 | 0.009 | 0.977 | 0.001 | 6 |
| Minus tail 35 | 1.721 | 2.730 | 0.023 | 0.985 | 0.000 | 6 |
| Minus tail 36 | 1.724 | 2.731 | 0.023 | 0.984 | 0.000 | 6 |
| Minus tail 37 | 1.745 | 2.784 | 0.024 | 0.983 | 0.000 | 6 |
| Minus tail 38 | 1.712 | 2.793 | 0.020 | 0.987 | 0.000 | 6 |
| Minus tail 39 | 1.709 | 2.767 | 0.020 | 0.986 | 0.000 | 6 |
| Minus tail 40 | 1.722 | 2.803 | 0.021 | 0.985 | 0.000 | 6 |
| Minus head 1 | 1.582 | 2.782 | 0.008 | 0.882 | 0.000 | 17 |
| Minus head 2 | 1.581 | 2.773 | 0.008 | 0.882 | 0.000 | 17 |
| Minus head 3 | 1.582 | 2.767 | 0.009 | 0.882 | 0.000 | 17 |
| Minus head 4 | 1.580 | 2.756 | 0.009 | 0.880 | 0.000 | 17 |
| Minus head 5 | 1.578 | 2.745 | 0.009 | 0.879 | 0.000 | 17 |
| Minus head 6 | 1.578 | 2.736 | 0.009 | 0.878 | 0.000 | 17 |
| Minus head 7 | 1.578 | 2.726 | 0.009 | 0.878 | 0.000 | 17 |
| Minus head 8 | 1.576 | 2.714 | 0.009 | 0.878 | 0.000 | 17 |
| Minus head 9 | 1.602 | 2.733 | 0.011 | 0.885 | 0.000 | 17 |
| Minus head 10 | 1.623 | 2.738 | 0.012 | 0.891 | 0.000 | 18 |
| Minus head 11 | 1.634 | 2.821 | 0.012 | 0.886 | 0.000 | 19 |
| Minus head 12 | 1.656 | 2.695 | 0.016 | 0.900 | 0.000 | 18 |
| Minus head 13 | 1.657 | 2.685 | 0.017 | 0.901 | 0.000 | 18 |
| Minus head 14 | 1.658 | 2.677 | 0.017 | 0.901 | 0.000 | 18 |
| Minus head 15 | 1.660 | 2.674 | 0.017 | 0.900 | 0.000 | 18 |
| Minus head 16 | 1.661 | 2.665 | 0.018 | 0.900 | 0.000 | 18 |
| Minus head 17 | 1.661 | 2.652 | 0.018 | 0.900 | 0.000 | 18 |
| Minus head 18 | 1.660 | 2.637 | 0.018 | 0.900 | 0.000 | 18 |
| Minus head 19 | 1.658 | 2.624 | 0.019 | 0.897 | 0.000 | 18 |
| Minus head 20 | 1.661 | 2.621 | 0.019 | 0.896 | 0.000 | 18 |
| Minus head 21 | 1.661 | 2.612 | 0.019 | 0.895 | 0.000 | 18 |
| Minus head 22 | 1.664 | 2.609 | 0.020 | 0.894 | 0.000 | 18 |
| Minus head 23 | 1.675 | 2.631 | 0.020 | 0.895 | 0.000 | 18 |
| Minus head 24 | 1.675 | 2.622 | 0.021 | 0.892 | 0.000 | 18 |
| Minus head 25 | 1.654 | 2.596 | 0.019 | 0.895 | 0.000 | 18 |
| Minus head 26 | 1.659 | 2.606 | 0.019 | 0.895 | 0.000 | 18 |
| Minus head 27 | 1.660 | 2.594 | 0.020 | 0.896 | 0.000 | 18 |
| Minus head 28 | 1.664 | 2.591 | 0.020 | 0.894 | 0.000 | 18 |
| Minus head 29 | 1.670 | 2.595 | 0.021 | 0.892 | 0.000 | 18 |
| Minus head 30 | 1.681 | 2.577 | 0.023 | 0.890 | 0.000 | 18 |
| Minus head 31 | 1.674 | 2.547 | 0.023 | 0.886 | 0.000 | 18 |
| Minus head 32 | 1.681 | 2.549 | 0.024 | 0.885 | 0.000 | 18 |
| Minus head 33 | 1.683 | 2.538 | 0.024 | 0.885 | 0.000 | 18 |
| Minus head 34 | 1.683 | 2.525 | 0.025 | 0.883 | 0.000 | 18 |
| Minus head 35 | 1.682 | 2.506 | 0.025 | 0.882 | 0.000 | 18 |
| Minus head 36 | 1.705 | 2.448 | 0.031 | 0.884 | 0.000 | 18 |
| Minus head 37 | 1.705 | 2.428 | 0.032 | 0.883 | 0.000 | 18 |
| Minus head 38 | 1.704 | 2.407 | 0.033 | 0.881 | 0.000 | 18 |
| Minus head 39 | 1.703 | 2.385 | 0.034 | 0.882 | 0.000 | 18 |
| Minus head 40 | 1.705 | 2.377 | 0.034 | 0.883 | 0.000 | 18 |

**Table S8.** The randomization (permutation) test for the TPL stability model parameters between COVID-19 and SARS parameters

| Treatments | Treatment A | Treatment B | TPL Parameter | p-value Cumulative | p-value Incremental |
|---|---|---|---|---|---|
| COVID-19 | Whole world | Except China | $b$ | 0.989 | 0.848 |
| | | | $\ln(a)$ | 0.977 | 0.991 |
| | | | $M_0$ | 0.984 | 0.889 |
| SARS | Whole world | Except China | $b$ | 0.688 | 0.964 |
| | | | $\ln(a)$ | 0.703 | 0.974 |
| | | | $M_0$ | 0.830 | 0.969 |
| Whole world | SARS | COVID-19 | $b$ | 0.874 | 0.102 |
| | | | $\ln(a)$ | 0.028 | 0.000 |
| | | | $M_0$ | 0.002 | 0.002 |
| Except China | SARS | COVID-19 | $b$ | 0.791 | 0.157 |
| | | | $\ln(a)$ | 0.206 | 0.000 |
| | | | $M_0$ | 0.002 | 0.001 |